\documentclass[rnote,traditabstract]{aa}

\usepackage[fleqn]{amsmath}
\usepackage{amssymb}
\usepackage{natbib}
\usepackage{txfonts}
\usepackage{graphicx}
\usepackage{lscape}
\usepackage{url}
\usepackage[version=3]{mhchem}
\usepackage{subfigure}

\begin{document}

\title{Photodesorption of \ce{H2O}, \ce{HDO}, and \ce{D2O} ice and 
its impact on fractionation 
\thanks{Appendices are available in electronic form at \protect\url{http://www.aanda.org}}\fnmsep
\thanks{Compiled simulation data and raw data are only available at the CDS via anonymous ftp 
to \protect\url{cdsarc.u-strasbg.fr (130.79.128.5)} or via \protect\url{http://cdsarc.u-strasbg.fr/viz-bin/qcat?J/A+A/vol/page}}}
\titlerunning{Photodesorption of \ce{H2O}, \ce{HDO}, and \ce{D2O} ice}

\author{Carina Arasa \inst{\ref{inst1},\ref{inst2}}
\and  Jesper Koning \inst{\ref{inst1}}
\and Geert-Jan Kroes \inst{\ref{inst1}}
\and Catherine Walsh \inst{\ref{inst2}}
\and Ewine F. van Dishoeck \inst{\ref{inst2},\ref{inst3}}} 
\authorrunning{Carina Arasa et al.}

\institute{Gorlaeus Laboratories, Leiden Institute of Chemistry, Leiden University, 
P. O. Box 9502, 2300 RA Leiden, The Netherlands \label{inst1} 
\and Leiden Observatory, Leiden University, P. O. Box 9513, 2300 RA Leiden, The Netherlands \label{inst2} 
\and Max Planck Institute for Extraterrestrial Physics, Giessenbachstrasse 1, 85748 Garching, Germany \label{inst3}\\
\email{ewine@strw.leidenuniv.nl}}

\date{Received 17 September 2013 / Accepted 16 December 2014}

\abstract {The \ce{HDO}/\ce{H2O} ratio measured in interstellar gas is 
often used to draw conclusions on the formation and evolution of 
water in star-forming regions and, by comparison with cometary data, 
on the origin of water on Earth.  In cold cores and in the outer 
regions of protoplanetary disks, an important source of gas-phase 
water comes from photodesorption of water ice.  This research 
note presents fitting formulae for implementation in astrochemical 
models using previously computed photodesorption efficiencies for 
all water ice isotopologues obtained with classical molecular 
dynamics simulations.   The results are used to investigate to 
what extent the gas-phase \ce{HDO}/\ce{H2O} ratio reflects that 
present in the ice or whether fractionation can occur during the 
photodesorption process.  Probabilities for the top four 
monolayers are presented for photodesorption of X (X=H,D) atoms, 
OX radicals, and \ce{X2O} and \ce{HDO} molecules following 
photodissociation of \ce{H2O}, \ce{D2O}, and \ce{HDO} in \ce{H2O} 
amorphous ice at ice temperatures from 10--100 K.  Significant 
isotope effects are found for all possible products: (1) H atom 
photodesorption probabilities from \ce{H2O} ice are larger than 
those for D atom photodesorption from \ce{D2O} ice by a factor of 
1.1; the ratio of H and D photodesorbed upon \ce{HDO} 
photodissociation is a factor of 2. This process will enrich the ice 
in deuterium atoms over time; (2) the OD/OH photodesorption ratio 
upon \ce{D2O} and \ce{H2O} photodissociation is on average a factor 
of 2, but the OD/OH photodesorption ratio upon \ce{HDO} 
photodissociation is almost constant at unity for all ice 
temperatures; (3) D atoms are more effective in kicking out 
neighbouring water molecules than H atoms. 
However, the ratio of the photodesorbed HDO and \ce{H2O} molecules 
is equal to the \ce{HDO}/\ce{H2O} ratio in the ice, therefore, there is no 
isotope fractionation when \ce{HDO} and \ce{H2O} photodesorb from the ice.  
Nevertheless, the enrichment of the ice in D atoms due to photodesorption can over 
time lead to an enhanced \ce{HDO}/\ce{H2O} ratio in the ice, and, 
when photodesorbed, also in the gas. The extent to which the 
ortho/para ratio of \ce{H2O} can be modified by the photodesorption 
process is discussed briefly as well.}


\maketitle

\section{Introduction}
\label{introduction}

The abundance of water in interstellar clouds, its partitioning
between gas and ice, and its evolution from collapsing cores to
protoplanetary disks are topics that are directly relevant for the
formation of planets and their atmospheres.  Astronomers can study
the water trail starting either from the very first stages of
dense molecular clouds even before a star is born
\citep{Whittet88,Boogert08,Caselli12} or working backwards from the
oceans on our own planet Earth, trying to link its composition to that
of icy solar system bodies (e.g., \citealt{Nuth08}, \citealt{Mumma11}, and 
see summary in \citealt{vanDishoeck14}).  
In both approaches, the \ce{HDO}/\ce{H2O} ratio is thought to be a good indicator of the
relevant processes involving water.  Specifically, the fact that the
\ce{HDO}/\ce{H2O} abundance of 3.1$\times 10^{-4}$ in Earth's oceans
\citep{Laeter03} is a factor of 2 lower than that found in most comets
\citep{Villanueva09} has been used as an argument that most water on
Earth originates from impact of asteroids rather than comets
\citep{Raymond04}\footnotemark.  Recent observations with the 
{\it Herschel Space Observatory} have found at least two comets 
for which the \ce{HDO}/\ce{H2O} ratio is nearly the same as that 
in Earth's oceans, demonstrating that there is a population of 
icy bodies with similar values \citep{Hartogh11,Bockelee12}. 
However, `hot off the press' data from the {\it Rosetta} mission to comet 
67~P/Churyumov-Gerasimenko now suggest that the HDO/\ce{H2O} in Jupiter 
family objects has a wider range than heretofore assumed triggering further 
debate on the origin of Earth's water \citep{Altwegg14}.
Hence, there is increased interest in understanding what 
processes affect the \ce{HDO}/\ce{H2O} ratio from core to disk 
to icy planetary bodies. 

\footnotetext{The measured \ce{HDO}/\ce{H2O} ratio is 2~$\times$ (D/H) in water; the 
latter values are commonly used in the literature.}

Physical-chemical models of the collapse of interstellar clouds and
disk formation suggest that the bulk of water is formed on grains in
the dense pre-collapse stage and enters the disk as ice that is
largely unaltered since its formation
\citep[e.g.,][]{Aikawa99,Visser09}.  
If so, the \ce{HDO}/\ce{H2O}
ratio in the disk (and eventually in the planetesimals and planets
that formed from it) should reflect the ice values found in the
interstellar cloud from which the star formed.  
Recent simulations of deuterium chemistry in protoplanetary disks 
show that, in the absence of strong vertical mixing, 
a high initial \ce{HDO}/\ce{H2O} ratio is required in the ice 
to reproduce the values seen in solar system bodies 
\citep{Furuya13,Albertsson14,Cleeves14}.
In cold clouds prior
to star formation and in the cold outer envelopes of low-mass
protostars ($T<20$~K), high ratios of deuterated species up to a few
$\times~10^{-2}$ have been found for many molecules including water
(e.g., \citealt{vanDishoeck95}, \citealt{Ceccarelli98}, \citealt{Bacmann03}, 
\citealt{Parise04}, \citealt{Parise12} and see review by \citealt{Ceccarelli14}).  
Although the fractionation of \ce{HDO} is generally not as high 
as that of other molecules, values of HDO/\ce{H2O} up to a few \% have 
been inferred in cold gas \citep{Liu11,Coutens12}, higher than
those found in comets.  They are also high compared with the upper
limits on \ce{HDO}/\ce{H2O} measured directly in interstellar ices of
$< (1-5) \times 10^{-3}$ \citep{Dartois03,Parise03}.  Moreover, the
\ce{HDO}/\ce{H2O} ratio in warm gas ($>100$~K) near protostars
typically has lower values of $\sim 10^{-3}$, within a factor of
a few of those of comets
\citep[e.g.,][]{Gensheimer96,vanderTak06,Jorgensen10,Persson14,Emprechtinger13,Neill13,Coutens14a,Coutens14b}.
These differences raise the question whether the high
\ce{HDO}/\ce{H2O} values measured in cold gas reflect the
\ce{HDO}/\ce{H2O} value in the ice or whether they have been altered
by desorption processes.

In cold clouds, the bulk of the water is present as ice with only a
small fraction of water in the gas, typically $10^{-4}$ of that of ice
\citep[e.g.,][]{Boonman03,Caselli12}.  The water ice is formed in an
early stage of cloud formation once the density and extinction reach a
threshold \citep{Cuppen07}.  Temperatures in these regions are too low
for thermal sublimation of ice, so the observed cold gas-phase water 
must come from non-thermal desorption processes of water ice 
with photodesorption a leading candidate \citep{Hollenbach09,Oberg09}.  
Indeed, both in pre-stellar cores
\citep{Caselli12}, in the outer parts of protostellar envelopes
\citep{Mottram13,Schmalzl14}, and in the cold outer parts of disks
\citep{Dominik05,Hogerheijde11}, the small observed gas-phase water
amount is consistent with being due just to the photodesorption of
water ice.  The required UV photons are provided by the general
interstellar radiation field at the outside of the cloud or by the
star in the case of a protoplanetary disk, and by the interaction of
cosmic rays with \ce{H2} deeper inside the cloud.  
Other non-thermal processes such as chemical desorption are poorly 
quantified but unlikely to be significant for thick layers of water 
ice \citep{Dulieu13,Minissale14}.  
Chemical models have so far considered mostly deuterium 
fractionation processes in the gas phase or on the grains, but not 
during the (photo)desorption process 
(e.g., \citealt{Tielens83}, \citealt{Aikawa99}, \citealt{Roberts03}, 
\citealt{Cazaux11}, \citealt{Du12}, \citealt{Aikawa12}, \citealt{Taquet13} and 
see summary in \citealt{vanDishoeck13}).

In this paper, we use previously computed photodesorption probabilities to 
investigate the extent to which the photodesorption
process can modify the gas-phase \ce{HDO}/\ce{H2O} ratio compared with
that of the ices.  Experiments have demonstrated that the water
photodesorption yield is typically $10^{-3}$ per incident photon, but
those data did not have the accuracy to establish differences between
\ce{H2O} and its isotopologues \citep{Westley95,Oberg09}.
Photodesorption can also be quantified by molecular dynamics
simulations.  In a series of papers, both the photodesorption
mechanisms and yields have been investigated for various water
isotopologues and for different ice temperatures
\citep{Andersson06,Andersson08,Arasa10,Arasa11,Koning13}.  The
incident UV photon is absorbed by a water molecule in the ice, which
dissociates into H~+~OH, with both fragments having excess energy.
The outcome depends on the ice layer in which the UV photon is
absorbed; only the top layers actively participate in desorption.  
Intact water molecules are released to the gas with a
yield of $\sim 5 \times 10^{-4}$ per incident photon through two
processes: (i) recombination of the H~+~OH fragments followed by
escape of the energetic newly formed water molecule; and (ii) kick-out
of a neighbouring water molecule by the energetic H atom produced by
photodissociation. 
In contrast with molecules such as CO, the excited water molecule does 
not directly desorb since all UV absorptions immediately lead to dissociation 
of the molecule.
\citet{Koning13} have investigated the yields for the various
processes using all combinations of water and its isotopologues, i.e.,
\ce{HDO} and \ce{D2O} in \ce{H2O} ice but also \ce{H2O} in \ce{D2O}
ice.  Isotope-selective effects are found for the various combinations
because of the different masses of the fragments.  For example, if a D
atom is created upon photodissociation, it is more effective in
kicking out a neighbouring molecule than an H atom because of more
efficient momentum transfer.

We here summarise the calculated photodesorption efficiencies as a
function of ice layer and ice temperature for the
astronomically relevant cases of \ce{H2O}, \ce{HDO}, and \ce{D2O} in
\ce{H2O} ice, and we provide convenient fitting formulae with depth
into the ice for use in astrochemical models.  Subsequently the
isotope-selective effects are quantified.  Specifically, the more
effective desorption of atomic H compared with atomic D will result in
enrichment of D in the ice.  
Furthermore, the extent to which the photodesorption process can
affect the ortho/para ratio of \ce{H2O} in the gas is briefly
discussed.  Sect.~\ref{methods} briefly summarises the computational
methods that were used in previous papers.
Sect.~\ref{results} lists the results for the isotope selective
processes for each ice layer at ice temperatures from 10--90~K.
Sect.~\ref{conclusions} summarises the conclusions and astrophysical
implications.

\section{Methods} 
\label{methods}

Our methods have been explained in detail in our previous studies
\citep{Andersson06,Andersson08,Arasa10,Arasa11,Koning13}, and are
based on classical Molecular Dynamics (MD) methods \citep{Allen87}, in
which the atoms and molecules in water ice move according to Newton's
equations based on analytical potentials of the interactions.  In
brief, an amorphous ice consisting of 480 water molecules is built for
a certain ice temperature, and one water molecule (\ce{X2O} (X=H or D)
or \ce{HDO}) is then randomly selected to be dissociated by UV
radiation.  Our previous studies show that only the photoexcited
molecules that are initially located in the top four monolayers lead
to photodesorption processes.

\begin{figure*}[tbp]
\centering
\includegraphics[width=\textwidth]{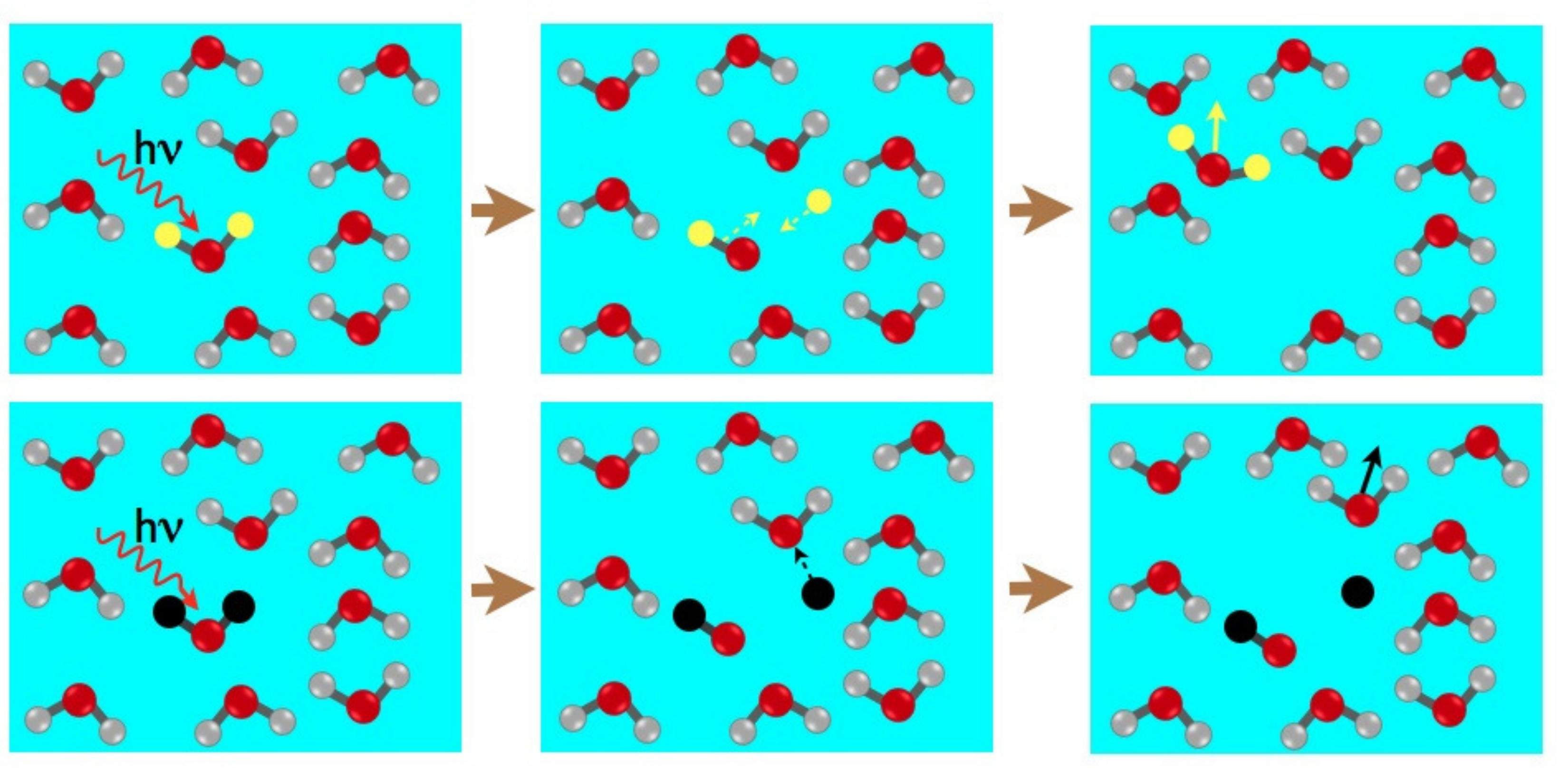}
\caption{Top: A \ce{H2O} molecule (with H atoms in yellow) 
surrounded by water molecules absorbs an UV photon and dissociates into H and OH. 
H and OH recombine to form \ce{H2O} that eventually desorbs from the ice surface via the direct mechanism.  
Bottom: A \ce{D2O} molecule (with D atoms in black) surrounded by water molecules absorbs an UV photon  
and dissociates into D and OD.  
The heavier D atom transfers its momentum to one of the surrounding \ce{H2O} molecules that desorbs from the ice 
via the ``kick-out" mechanism.}
\label{fig:AAcartoon}
\end{figure*}

The dynamics of the photodissociation fragments are subsequently
followed in the ice over a timescale of a few picoseconds until one of
six possible outcomes is reached (see Fig.~\ref{fig:AAcartoon}): (1) X
atom desorption while OX stays trapped in the ice, (2) OX radical
desorption while X stays trapped in the ice, (3) both X and OX desorb,
(4) trapping of both X and OX photofragments, (5) recombination of the
X atom and the OX radical to form \ce{X2O} which in the end desorbs,
or (6) stays trapped in the ice.  When \ce{HDO} is dissociated the
outcome channels are the same, but the recombination of the
photofragments leads to \ce{HDO} that either desorbs (5), or stays
trapped in the ice (6).  
Moreover, there is a parallel outcome to any of the 6 processes by
which one of the surrounding \ce{H2O} molecules from the ice can also
desorb, initiated by the same UV photon.  This mechanism is called
``kick-out'' \citep{Andersson06,Andersson08}, and it takes place when
the energetic X atom produced by photodissociation kicks a surrounding
molecule from the ice by transfer of its momentum, and eventually the
kicked out \ce{H2O} molecule desorbs from the ice.  

The calculations are repeated $\ge$~6000 times per species and  
per monolayer so that the statistics (i.e., probabilities) of the different 
outcomes are determined as a function of the initial position of the molecule in the ice. 
The water molecules are excited with UV radiation in the 7.5--9.5 eV (1650-1300
\AA) range, corresponding to the first electronic absorption band of
water ice.  Throughout the paper, $^*$ indicates the molecule that is
photodissociated in the ice.  For the case of \ce{HDO}$^*$ there are
two possible outcomes: HOD$^*$ indicates photodissociation to H~+~OD
whereas DOH$^*$ denotes dissociation to D~+~OH; \ce{HDO} and
\ce{HDO}$^*$ denote the generic case.

In our previous studies \citep{Arasa10, Arasa11} we calculated the
total photodesorption yields of OX and \ce{X2O} per incident photon,
and compared the results with the available experimental data
\citep{Oberg09}.  For our model \ce{D2O} ice, the photodesorption
yield agrees well with the experimental photodesorption yield at low
ice temperatures within the $\sim$~60\%~experimental uncertainties
\citep{Arasa11}.  At higher ice temperatures, the experimental
photodesorption yield is larger than the calculated value due to long
timescale thermal effects that cannot be considered in our
simulations.  The experimental OX/\ce{H2O} ratio is however on average
a factor of 2 to 3 lower than the calculated value for \ce{H2O} ice
(see Table II in \citealt{Arasa10}), and \ce{D2O} ice, respectively.
In a different kind of experiment, \citet{Yabushita09} measured the
final translational and rotational energies of the kicked out \ce{H2O}
molecules for $\upsilon$~=~0 at $T_\mathrm{ice }$~=~90~K; their values
match well with our calculated ones at $T_\mathrm{ice }$~=~90~K and
provide proof for the importance of the kick-out mechanism
\citep{Arasa10}.  These comparisons between models and data provide
confidence in the accuracy of our simulations. For isotope selective
processes, only relative trends are considered, which should have less
uncertainty than the absolute values.

\section{Results and discussion}
\label{results}

In this section we summarize the probabilities of all the
photodesorption events after \ce{H2O}$^*$, \ce{HDO}$^*$ and
\ce{D2O}$^*$ photodissociation in amorphous \ce{H2O} ice at relevant
temperatures, $T_\mathrm{ice }$~=~10 and 90~K.  
The supplementary material contains the probabilities at 
$T_\mathrm{ice }$~=~20, 30, and 60~K for the top four monolayers (ML) 
of the ice surface.  
These results have been reported and extensively discussed elsewhere \citep{Koning13}, 
but they are presented here in tabulated form for each of the top four monolayers 
so that the data can be used in astrochemical models.  
In addition, we provide fitting formulae to these tables in Table~\ref{TableFits} 
(see Appendix B for more details).  
All tables can be downloaded from \url{http://www.strw.leidenuniv.nl/}$\sim$\url{ewine/photo}.

A total of 6000 trajectories per monolayer have been simulated for photoexcitation 
of \ce{H2O} and \ce{D2O}.  For photoexcitation of HDO, $\ge$~6000 trajectories were 
run per monolayer. 
The photodesorption probabilities are calculated for each event as
$p_i = {N_i}/{N_\mathrm{total}}$, $i$~=~photodesorption event,
$N_i$~=~number of trajectories that lead to the photodesorption event
$i$, $N_\mathrm{total}$~=~total number of trajectories per monolayer
$j$, the average is taken over the top four monolayers 
(e.g., $\langle p_{i} \rangle = \sum_{j=1}^{4}p_{i}^{j}/4$).
All the probabilities are given per {\it absorbed} (rather than incident) UV
photon.  The averaged photodesorption probabilities calculated per
incident UV photon are on average a factor of $\sim$~0.03 smaller than
those given per absorbed UV photon for processes (1), (2), (3) and (5)
\citep[see numbering of outcomes in Sect.~2 of this paper and the description in][]{Arasa10,Arasa11}.

Above 100~K, thermal desorption takes over as the main desorption
mechanism in interstellar space, so our results are not relevant for
the interpretation of data on warm water.

\begin{table}[tbp]
\centering
\caption{Fitting functions and best-fit values of the parameters $a$, $b$, $c$, and ML, ML being the number of the monolayer.}
\begin{tabular}{cccc}
\hline\hline
Species & $a$ & $b$ & $c$ \\
\hline
&\multicolumn{3}{c}{X$_\mathrm{des}$ + OY$_\mathrm{trapped}$} \\
&\multicolumn{3}{c}{$P(\ce{ML}) = (a\ce{ML})\exp{(-b\ce{ML})}$} \\
\hline
\ce{H2O} & 2.02 & 0.858 & $\cdots$ \\
\ce{D2O} & 2.06 & 0.906 & $\cdots$ \\
\ce{HOD} & 2.04 & 0.867 & $\cdots$ \\
\ce{DOH} & 2.17 & 0.942 & $\cdots$ \\
\hline
&\multicolumn{3}{c}{X$_\mathrm{trapped}$ + OY$_\mathrm{des}$} \\
&\multicolumn{3}{c}{$P(\ce{ML}) = (a\ce{ML})\exp{(-b(\ce{ML}-c)^2)}$} \\
\hline
\ce{H2O} & 4.28    & 0.0485 & -10.9 \\
\ce{D2O} & 0.0141  & 0.377  & 0.264 \\
\ce{HOD} & 0.00359 & 0.635  & 0.990 \\
\ce{DOH} & 0.0114  & 0.450  & 0.751 \\
\hline
&\multicolumn{3}{c}{X$_\mathrm{des}$ + OY$_\mathrm{des}$} \\
&\multicolumn{3}{c}{$P(\ce{ML}) = (a\ce{ML})\exp{(-b(\ce{ML}-c)^2)}$} \\
\hline
\ce{H2O} & 0.0710 & 0.210 & -1.77  \\
\ce{D2O} & 0.0638 & 0.554 & 0.269  \\
\ce{HOD} & 0.0463 & 0.383 & -0.383 \\
\ce{DOH} & 0.0659 & 0.484 & 0.0538 \\
\hline
&\multicolumn{3}{c}{X$_\mathrm{trapped}$ + OY$_\mathrm{trapped}$} \\
&\multicolumn{3}{c}{$P(\ce{ML}) = a(1 - \exp{(-b\ce{ML}^c)})$} \\
\hline
\ce{H2O} & 0.516 & 0.149 & 1.48 \\
\ce{D2O} & 0.458 & 0.163 & 1.57 \\
\ce{HOD} & 0.595 & 0.125 & 1.60 \\
\ce{DOH} & 0.574 & 0.111 & 1.89 \\
\hline
&\multicolumn{3}{c}{XYO$_\mathrm{direct}$} \\
&\multicolumn{3}{c}{$P(\ce{ML}) = (a\ce{ML})\exp{(-b(\ce{ML}-c)^2)}$} \\
\hline
\ce{H2O} & 0.390   & 0.0819 & -6.04  \\
\ce{D2O} & 0.00523 & 0.470  & 0.967 \\
\ce{HOD} & 0.00428 & 0.604  & 1.12  \\
\ce{DOH} & 0.00494 & 0.570  & 1.11  \\
\hline
&\multicolumn{3}{c}{\ce{H2O}$_\mathrm{kicked}$}  \\
&\multicolumn{3}{c}{$P(\ce{ML}) = (a\ce{ML})\exp{(-b(\ce{ML}-c)^2)}$} \\
\hline
\ce{H2O} & 0.00320 & 0.210 & 0.882 \\
\ce{D2O} & 0.00865 & 0.452 & 1.83  \\
\ce{HOD} & 0.00465 & 0.193 & 0.419 \\
\ce{DOH} & 0.00949 & 0.180 & 0.519 \\
\hline
\end{tabular}
\tablefoot{The species XOY denotes photoexcitation of XOY leading to dissociation into X + OY.  
The units of the parameters, $a$, $b$, and $c$, depend on the 
fitting function which itself describes a probability and thus has no physical units.  
For outcome (1), X$_\mathrm{des}$ + OY$_\mathrm{trapped}$, $a$ and $b$ have units of ML$^{-1}$, where ML is monolayer number.  
For outcome (4), X$_\mathrm{trapped}$ + OY$_\mathrm{trapped}$, $a$ has no units and $b$ has units of ML$^{-c}$.  
For all other outcomes, $a$ has units of ML$^{-1}$, $b$ has units of ML$^{-2}$, and $c$ has units of ML.}
\label{TableFits}
\end{table}

\subsection{X (X=H,D) atom photodesorption probabilities} 
\label{Xprobabilities}

The X atom photodesorption event is the dominant process when the
\ce{X2O} photodissociated molecule is located in the top four
monolayers of the ice surface \citep[see Table~\ref{Table1} and][]{Andersson06,Andersson08}.  
The X atom photodesorption probabilities are $>$~90\% in the top monolayer (ML1) and decrease
with depth to $\sim$~30\% in ML4 because in the top MLs the X atoms
can easily find their way to escape from the surface, whereas deeper
in the ice there are other molecules that prevent their desorption.

In general, H atom photodesorption probabilities are somewhat larger following
\ce{H2O} dissociation than those for D atom photodesorption after
\ce{D2O} photodissociation in \ce{H2O} ice, especially if initially
the molecules are located in the third and fourth monolayers. 
The same trend is observed after HOD$^*$/DOH$^*$ photodissociation in
\ce{H2O} ice: H atom photodesorption probabilities are larger than
those for the D atom. 
This is because of isotope mass effects: the H atom is lighter than the D atom 
and therefore transfers less energy to nearby water molecules, 
so that it can travel more easily from
the bottom to the top of the surface and eventually desorb to the gas
phase \citep{Arasa11, Koning13}.

The average H/D photodesorption ratios following \ce{H2O}$^*$ and
\ce{D2O}$^*$ photodissociation and the average H/D ratios following
\ce{HDO}$^*$ photodissociation over the top four MLs are summarized in
Table~\ref{table:dh-ratio} versus ice temperature. 
The photodesorption ratio $\frac{\ce{H}_\mathrm{des}}{\ce{D}_\mathrm{des}}$ after \ce{HDO}
photodissociation have been calculated (as was done for the photodesorption 
$\frac{\ce{OD}_\mathrm{des}}{\ce{OH}_\mathrm{des}}$ in Table~IV in \citealt{Koning13}) 
taking into account the probabilities in each ML, $i$,
where $P_{\ce{D}_\mathrm{des}}(i,T_\mathrm{ice})$ is the D photodesorption
probability after DOH$^*$ photodissociation into D and OH, 
and $P_{\ce{H}_\mathrm{des}}(i,T_\mathrm{ice})$ is the H photodesorption probability after
HOD$^*$ photodissociation into H and OD), and the branching ratios
($\frac{\ce{H}+\ce{OD}}{\ce{D}+\ce{OH}}$) 
in the ice in each ML, $i$, and temperature ($\beta$($i$, $T_\mathrm{ice}$)).  
The effect is typically a factor of 1.1 when comparing  \ce{H2O} and  \ce{D2O}, 
and when comparing HOD in \ce{H2O} and DOH in \ce{H2O}.
 
For \ce{HDO}$^*$ photodissociation, there is another effect at work, namely
that the probability of dissociation into H~+~OD is larger than that
into D~+~OH, a process that is well known in the gas phase. 
For gas-phase photodissociation, the OD/OH branching ratio is about 3.1
\citep{vanHarrevelt12}. 
In the ice, this effect is less, but still at the factor of two level.  
The \ce{HDO}$^*$ photodissociation branching ratios
($\frac{\ce{H}+\ce{OD}}{\ce{D}+\ce{OH}}$) in the ice in each ML, $i$, 
and temperature ($\beta$($i$, $T_\mathrm{ice}$)) were calculated and listed in Table~IV in
\citet{Koning13}.  
The average branching ratio taken over all ice temperatures is about 2.2 in favor of H~+~OD. 
Thus the branching ratio upon \ce{HDO} photodissociation in the ice reinforces the H/D desorption
ratio leading to a ratio more than a factor of two larger than the
ratio of H and D present in the ice (calculated from Tables~\ref{Table1} and \ref{Table4}). 
Therefore, all photons that arrive in the ice and are absorbed in
the top layers of the ice surface lead to an enrichment in
D atoms relative to H atoms in the ice mantles of dust particles.

\begin{table*}[tbp]
\centering
\caption{X atom photodesorption probabilities at $T_\mathrm{ice}$~=~10~K (top) and 90~K (bottom) resulting from photoexcitation 
of a \ce{X2O} (X=H,D) or XOY (HOD or DOH) molecule present in a specific monolayer of \ce{H2O} ice.} 
\begin{tabular}{ccccc} 
\hline\hline
ML        & H$_\mathrm{des}$/\ce{H2O}$^*$        & D$_\mathrm{des}$/\ce{D2O}$^*$       & H$_\mathrm{des}$/HOD$^*$            & D$_\mathrm{des}$/DOH$^*$           \\ \hline
\multicolumn{5}{c}{$T_\mathrm{ice}$~=~10~K} \\ \hline
1         & 0.920 $\pm$ 3.9$\times$10$^{-3}$ & 0.900 $\pm$ 4.0$\times$10$^{-3}$  & 0.911 $\pm$ 2.8$\times$10$^{-3}$  & 0.912 $\pm$ 3.5$\times$10$^{-3}$ \\
2         & 0.700 $\pm$ 5.8$\times$10$^{-3}$ & 0.681 $\pm$ 6.0$\times$10$^{-3}$  & 0.751 $\pm$ 4.3$\times$10$^{-3}$  & 0.710 $\pm$ 5.6$\times$10$^{-3}$ \\  
3         & 0.510 $\pm$ 6.0$\times$10$^{-3}$ & 0.488 $\pm$ 6.3$\times$10$^{-3}$  & 0.521 $\pm$ 4.9$\times$10$^{-3}$  & 0.472 $\pm$ 6.0$\times$10$^{-3}$ \\
4         & 0.340 $\pm$ 5.2$\times$10$^{-3}$ & 0.231 $\pm$ 5.4$\times$10$^{-3}$  & 0.259 $\pm$ 4.3$\times$10$^{-3}$  & 0.190 $\pm$ 4.7$\times$10$^{-3}$ \\ 
$\langle$MLs$\rangle$ & 0.617 $\pm$ 3.2$\times$10$^{-3}$ & 0.575 $\pm$ 3.2$\times$10$^{-3}$  & 0.610 $\pm$ 4.1$\times$10$^{-3}$  & 0.572 $\pm$ 5.0$\times$10$^{-3}$ \\ \hline
\multicolumn{5}{c}{$T_\mathrm{ice}$~=~90~K} \\ \hline
1         & 0.922 $\pm$ 3.9$\times$10$^{-3}$ & 0.891 $\pm$ 4.0$\times$10$^{-3}$  & 0.910 $\pm$ 3.4$\times$10$^{-3}$  & 0.910 $\pm$ 3.5$\times$10$^{-3}$ \\
2         & 0.763 $\pm$ 5.3$\times$10$^{-3}$ & 0.682 $\pm$ 6.0$\times$10$^{-3}$  & 0.660 $\pm$ 6.0$\times$10$^{-3}$  & 0.630 $\pm$ 6.0$\times$10$^{-3}$ \\  
3         & 0.425 $\pm$ 5.6$\times$10$^{-3}$ & 0.407 $\pm$ 6.3$\times$10$^{-3}$  & 0.460 $\pm$ 6.3$\times$10$^{-3}$  & 0.380 $\pm$ 6.0$\times$10$^{-3}$ \\
4         & 0.320 $\pm$ 3.4$\times$10$^{-3}$ & 0.231 $\pm$ 5.4$\times$10$^{-3}$  & 0.320 $\pm$ 6.0$\times$10$^{-3}$  & 0.250 $\pm$ 5.3$\times$10$^{-3}$ \\ 
$\langle$MLs$\rangle$ & 0.608 $\pm$ 3.2$\times$10$^{-3}$ & 0.553 $\pm$ 3.2$\times$10$^{-3}$  & 0.589 $\pm$ 5.5$\times$10$^{-3}$  & 0.543 $\pm$ 5.0$\times$10$^{-3}$ \\ \hline
\end{tabular}
\tablefoot{ML1 is the top monolayer and $\langle$MLs$\rangle$ denotes the average over the top four monolayers.}
\label{Table1}
\end{table*}

\begin{table}[tbp]
\centering
\caption{Average photodesorption ratio $\frac{\ce{H}_\mathrm{des}}{\ce{D}_\mathrm{des}}$ 
following \ce{H2O}$^*$, \ce{D2O}$^*$ and \ce{HDO}$^*$ photodissociation. 
For \ce{HDO}$^*$, the last column shows the result without considering the effect of branching in the photodissociation.}
\begin{tabular}{cccc} 
\hline
\hline
$T_\mathrm{ice}$ & \multicolumn{3}{c}{$\frac{\ce{H}_\mathrm{des}}{\ce{D}_\mathrm{des}}$} \\ 
(K)              & \ce{H2O}$^*$ and \ce{D2O}$^*$                       & \multicolumn{2}{c}{ \ce{HDO}$^*$} \\ \hline
10 &  1.07 $\pm$ 0.01 & 2.36 $\pm$ 0.01 & 1.01 $\pm$ 0.01  \\
20 &  1.06 $\pm$ 0.01 & 2.43 $\pm$ 0.02 & 1.19 $\pm$ 0.01 \\
30 &  1.06 $\pm$ 0.01 & 2.45 $\pm$ 0.02 & 1.08 $\pm$ 0.01  \\
60 &  1.06 $\pm$ 0.01 & 2.55 $\pm$ 0.02 & 1.11 $\pm$ 0.01  \\
90 &  1.10 $\pm$ 0.01 & 2.47 $\pm$ 0.02 & 1.08 $\pm$ 0.01  \\ \hline
\end{tabular}
\label{table:dh-ratio}
\end{table}

\subsection{OX (X=H,D) radical photodesorption probabilities} 
\label{OXprobabilities}

The second most important photodesorption event is OX photodesorption
with probabilities of typically a few \% in the top layers. 
In Table~\ref{Table2} the OX radical photodesorption probabilities for the top
four monolayers together with the average values taken over the top four
MLs are summarised at $T_\mathrm{ice}$~=~10~K and 90~K.  
It is seen that the photodesorption probabilities drop sharply from ML2 to ML3. 
This feature has been explained in detail previously
\citep{Andersson06,Andersson08,Koning13}. 
Although OX is quite mobile in the top ice layer (ML1), 
it loses much of its energy in lower layers to other \ce{H2O} molecules in the ice,
preventing desorption of OX.  

The OH photodesorption probabilities after DOH$^*$ photodissociation
and OD photodesorption probabilities after \ce{D2O}$^*$
photodissociation are larger than those for OD after HOD$^*$
photodissociation and OH after \ce{H2O}$^*$ in \ce{H2O} ice. 
This result is expected because the OD radicals have higher kinetic
energies than the OH radicals upon the initial \ce{D2O} and \ce{H2O}
photodissociation, respectively. 
In contrast, the OH radicals have higher kinetic energy than the 
OD radicals upon \ce{HDO}$^*$ photodissociation.
These findings follow from applying energy and momentum conservation
($p_\ce{OX}+p_\ce{X}=0$, and $E_\ce{OX}+E_\ce{X}=\Delta E$,
with $\Delta E$ being the initial available energy; 
$\Delta E = E_\mathrm{exc} - E_\mathrm{diss}(\ce{X2O})$, see \citealt{Arasa11}) . 
For \ce{X2O} photodissociation at the same excitation energy $E_\mathrm{exc}$ it is seen
that if  $E_\mathrm{OX} = \Delta E/(1 + m_\ce{OX}/{m_\ce{X}})$, 
the initial energy of OD should be a factor 1.8 larger than that for OH. 
In the case of X$^{(1)}$OX$^{(2)}$
photodissociation into OX$^{(2)}$ + X$^{(1)}$, and applying the same
conservation rules we can expect that the energy of the OX$^{(2)}$
radical formed upon DOX$^{(2)}$ photodissociation is about a factor of
two larger than the OX$^{(2)}$ radical formed upon HOX$^{(2)}$
photodissociation (see Table III in \citealt{Koning13}).  

\begin{figure}[tbp]
\centering
\includegraphics[width=0.5\textwidth,clip]{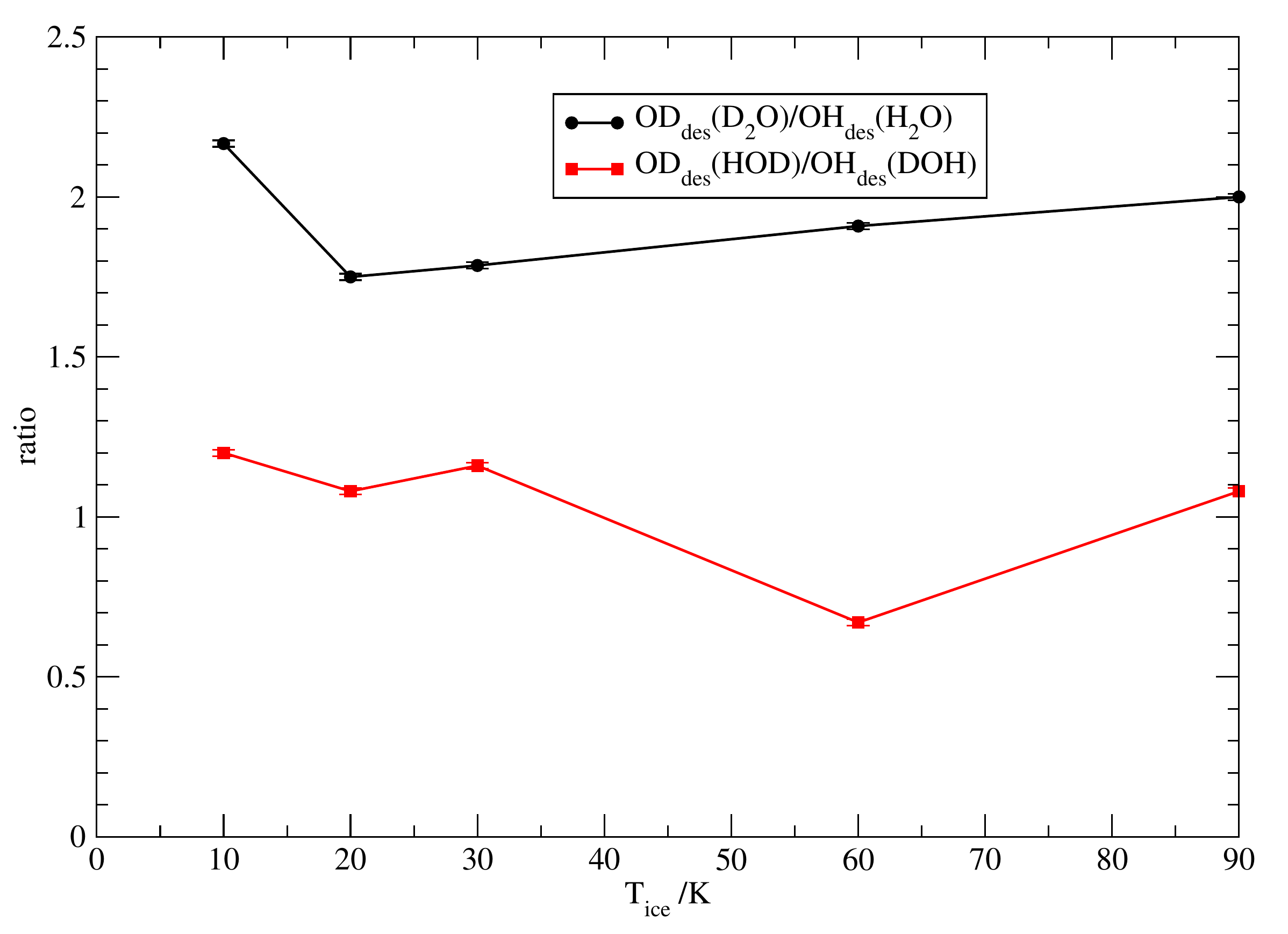}
\caption{OD$\rm{_{des}}$(\ce{D2O})/OH$_\mathrm{des}$(\ce{H2O}) ratios 
and OD$_\mathrm{des}$(HOD)/OH$\mathrm{_{des}}$(DOH) ratios averaged over the top four monolayers following \ce{HDO} dissociation in \ce{H2O} 
ice versus ice temperature $T_\mathrm{ice}$ \citep[based on][]{Koning13}.}
\label{fig:odoh-ratio}
\end{figure}

The average OD/OH photodesorption ratios following \ce{HDO}$^*$
photodissociation over the top four MLs are summarised in
Fig.~\ref{fig:odoh-ratio}.  The photodesorption ratio
OD$_\mathrm{des}$/OH$_\mathrm{des}$ after HOD$^*$ and DOH$^*$
photodissociation have been calculated taking account the
probabilities in each ML, $i$, where
$P_{\ce{OD}_\mathrm{des}}(i,T_\mathrm{ice})$ is the OD photodesorption
probability after HOD$^*$ photodissociation into H and OD, and
$P_{\ce{OH}_\mathrm{des}}(i,T_\mathrm{ice})$ is the OH photodesorption
probability after DOH$^*$ photodissociation into D and OH, and the
photodissociation branching ratios H+OD/D+OH in the ice in each ML,
$i$, and temperature ($\beta$($i$, $T_\mathrm{ice}$), see Table~IV in
\citealt{Koning13}).  These values are seen to be close to unity and
to be mostly constant with ice temperature.  The reason that the
OD$_\mathrm{des}$/OH$_\mathrm{des}$ ratios are close to unity for
\ce{HDO}$^*$is that a cancellation of the two effects occurs.  On the
one hand, the OH desorption is more efficient than that of OD, as
explained above.  However, this is offset by the OD/OH branching ratio
of about 2.2 in the \ce{HDO}$^*$ photodissociation, as discussed in
Sect.~\ref{Xprobabilities}.  As a result, the average photodesorption
ratio OD$_\mathrm{des}$/OH$_\mathrm{des}$ upon \ce{HDO}$^*$
photodissociation is about 1.0 with a standard deviation of 0.2 in the
ice \citep{Koning13}.  Note that this result is very different from
pure gas-phase chemistry, where OD is produced a factor of three more
rapidly than OH by photodissociation of gaseous HDO.

Fig.~\ref{fig:odoh-ratio} also contains the average
photodesorption ratios between OD desorption
following \ce{D2O} photodissociation in \ce{H2O} ice and OH desorption
following \ce{H2O} photodissociation in \ce{H2O} ice for all ice
temperatures considered. 
On average, the OD$_\mathrm{des}$(\ce{D2O})/OH$_\mathrm{des}$(\ce{H2O}) 
ratio is about 1.9.

\begin{table*}[tbp]
\centering
\caption{OX radical photodesorption probabilities at $T_\mathrm{ice}$~=~10~K (top) and 90~K (bottom) resulting from photoexcitation 
of a \ce{X2O} (X=H,D) or XOY (HOD or DOH) molecule present in a specific monolayer of \ce{H2O} ice.}
\begin{tabular} {ccccc} 
\hline
\hline
ML        & OH$_\mathrm{des}$/\ce{H2O}$^*$   & OD$_\mathrm{des}$/\ce{D2O}$^*$     & OD$_\mathrm{des}$/HOD$^*$         & OH$_\mathrm{des}$/DOH$^*$        \\ \hline
\multicolumn{5}{c}{$T_\mathrm{ice}$~=~10~K} \\ \hline
1         & 0.022 $\pm$ 1.9$\times$10$^{-3}$   & 0.065  $\pm$ 3.2$\times$10$^{-3}$  & 0.044 $\pm$ 2.0$\times$10$^{-3}$  & 0.071 $\pm$ 3.2$\times$10$^{-3}$ \\
2         & 0.021 $\pm$ 1.8$\times$10$^{-3}$   & 0.035  $\pm$ 2.8$\times$10$^{-3}$  & 0.014  $\pm$ 1.2$\times$10$^{-3}$ & 0.033 $\pm$ 2.2$\times$10$^{-3}$ \\  
3         & (3.0 $\pm$ 0.7)$\times$10$^{-3}$   & (4.0 $\pm$ 0.9)$\times$10$^{-3}$   & (9.6 $\pm$ 3.0)$\times$10$^{-4}$  & (5.1 $\pm$0.9)$\times$10$^{-3}$  \\
4         & 0.00                               & 0.00                               & (1.9 $\pm$ 1.3)$\times$10$^{-4}$  & 0.00                             \\ 
$\langle$MLs$\rangle$ & 0.012 $\pm$ 6.9$\times$10$^{-4}$   & 0.026 $\pm$ 1.0$\times$10$^{-3}$   & 0.015 $\pm$ 1.2$\times$10$^{-3}$  & 0.027 $\pm$ 2.0$\times$10$^{-3}$ \\  \hline
\multicolumn{5}{c}{$T_\mathrm{ice}$~=~90~K} \\ \hline
1         & 0.036 $\pm$ 1.8$\times$10$^{-3}$   & 0.065  $\pm$ 3.2$\times$10$^{-3}$  & 0.016 $\pm$ 1.6$\times$10$^{-3}$  & 0.039 $\pm$ 2.4$\times$10$^{-3}$ \\
2         & 0.025 $\pm$ 1.9$\times$10$^{-3}$   & 0.050 $\pm$ 2.8$\times$10$^{-3}$   & 0.023 $\pm$ 1.9$\times$10$^{-3}$  & 0.045 $\pm$ 2.6$\times$10$^{-3}$ \\  
3         & (2.0 $\pm$ 0.7)$\times$10$^{-3}$   & (5.0 $\pm$ 0.9)$\times$10$^{-3}$   & (3.8 $\pm$ 0.8)$\times$10$^{-3}$  & (7.4 $\pm$ 0.1)$\times$10$^{-3}$ \\
4         & 0.00                               & (3.3 $\pm$ 2.4)$\times$10$^{-4}$   & (1.6 $\pm$ 1.6)$\times$10$^{-4}$  & (3.0 $\pm$ 2.2)$\times$10$^{-4}$ \\ 
$\langle$MLs$\rangle$ & 0.015 $\pm$ 7.9$\times$10$^{-4}$   & 0.030 $\pm$ 4.0$\times$10$^{-4}$   & 0.011 $\pm$ 1.3$\times$10$^{-3}$  & 0.023 $\pm$ 2.0$\times$10$^{-3}$ \\ \hline
\end{tabular}
\tablefoot{ML1 is the top monolayer and $\langle$MLs$\rangle$ denotes the average over the top four monolayers.}
\label{Table2}
\end{table*}

\subsection{\ce{X2O} (X=H,D) and \ce{HDO} molecule photodesorption probabilities} 
\label{X2Oprobabilities}

\ce{X2O} photodesorption is the third most important photodesorption
event after \ce{H2O}$^*$ and \ce{D2O}$^*$ photodissociation with
probabilities of $\sim$~0.5\% in the top layers.  As discussed above,
two processes can occur: direct desorption of the recombined \ce{H2O}
or \ce{D2O} molecule or through the kick of the energetic X atom to a
surrounding \ce{H2O} molecule from the ice.  When HOD$^*$ dissociates
it can also lead to the recombination of the H and OD fragments that
eventually desorb as \ce{HDO}, or the H atom can kick out the
surrounding \ce{H2O} molecules from the ice.  Similarly, when DOH$^*$
dissociates the photofragments can also recombine and desorb, or the D
atom can kick out a \ce{H2O} molecule.  In Table~\ref{Table3} the
probabilities of both mechanisms for the top four MLs and its average
for the four different scenarios are reported at
$T_\mathrm{ice}$~=~10~K and 90~K.  Again, isotope effects are rather
noticeable if \ce{X2O} desorbs from the ice surface through the
kick-out mechanism \citep{Koning13} after the X atom transfers its
momentum to the surrounding \ce{H2O} molecules.  Since the D atom is
heavier than the H atom it can transfer its momentum more easily to a
surrounding \ce{H2O} molecule, leading to higher kick-out
photodesorption probabilities, and in consequence to higher total
\ce{X2O} and HOD photodesorption probabilities \citep{Arasa11,Koning13}. 
As demonstrated in Appendix C, this process does however not lead to fractionation 
of HDO/\ce{H2O} in the gas in \ce{H2O}-dominated ices, i.e., 
the gas-phase HDO/\ce{H2O} ratio produced by photodesorption reflects that originally in the ice.

\begin{table*}[tbp]
\footnotesize
\centering
\caption{Total \ce{X2O} (X=H,D) or XOY (HOD or DOH) photodesorption probabilities at $T_\mathrm{ice}$~=~10~K (top) and 90~K (bottom) per monolayer due to the direct and the 
kick-out mechanism for \ce{X2O} and \ce{XOY} photodissociation in \ce{H2O} ice.}
\begin{tabular}{ccccccccc} 
\hline
\hline
   & \multicolumn{2}{c}{\ce{H2O}$^*$} & \multicolumn{2}{c}{\ce{D2O}$^*$} & \multicolumn{2}{c}{HOD$^*$} & \multicolumn{2}{c}{DOH$^*$} \\
ML & \ce{H2O}$_\mathrm{direct}$ & \ce{H2O}$_\mathrm{kicked}$  & \ce{D2O}$_\mathrm{direct}$ & \ce{H2O}$_\mathrm{kicked}$ & HOD$_\mathrm{direct}$ 
   & \ce{H2O}$_\mathrm{kicked}$ & DOH$_\mathrm{direct}$ & \ce{H2O}$_\mathrm{kicked}$ \\ \hline
\multicolumn{9}{c}{$T_\mathrm{ice}$~=~10~K} \\ \hline
1         & 5.7 $\pm$ 1.1 & 0.5 $\pm$ 0.2 & 6.8 $\pm$ 1.1   & 12.8 $\pm$ 1.5  & 5.6 $\pm$ 0.7   & 2.2 $\pm$ 0.5  & 5.3 $\pm$ 0.9   &  2.9 $\pm$ 0.7 \\
2         & 7.4 $\pm$ 1.2 & 3.9 $\pm$ 1.3 & 9.2 $\pm$ 1.2   & 17.8 $\pm$ 1.7  & 5.1 $\pm$ 0.7   & 2.7 $\pm$ 0.5  & 7.0 $\pm$ 1.0   & 10.7 $\pm$ 1.3 \\  
3         & 1.7 $\pm$ 0.8 & 4.8 $\pm$ 1.1 & 1.0 $\pm$ 0.4   & 16.5 $\pm$ 1.6  & 2.9 $\pm$ 0.5   & 0.8 $\pm$ 0.3  & 1.6 $\pm$ 0.5   &  1.2 $\pm$ 0.4 \\
4         & 0.00          & 0.6 $\pm$ 0.3 & 0.3 $\pm$ 0.2   & 4.3 $\pm$ 0.9   & 0.1 $\pm$ 0.1   & 1.2 $\pm$ 0.3  & 0.00            &  0.9 $\pm$ 0.4 \\
$\langle$MLs$\rangle$ & 3.7 $\pm$ 0.4 & 2.5 $\pm$ 0.2 & 4.3 $\pm$ 0.4   & 6.6 $\pm$ 0.5   & 3.4 $\pm$ 0.6   & 1.7 $\pm$ 0.4  & 3.5 $\pm$ 0.7   &  3.9 $\pm$ 0.8 \\ \hline 
\multicolumn{9}{c}{$T_\mathrm{ice}$~=~90~K} \\ \hline
1         & 5.3 $\pm$ 1.1 & 1.4 $\pm$ 0.8  & 6.8 $\pm$ 0.1  & 12.8 $\pm$ 1.5   & 2.6 $\pm$ 0.6  &  5.4 $\pm$ 0.9  & 6.8 $\pm$ 0.1  & 15.6 $\pm$ 1.5 \\
2         & 9.6 $\pm$ 1.3 & 3.9 $\pm$ 0.9  & 9.2 $\pm$ 1.2  & 17.8 $\pm$ 1.7   & 6.7 $\pm$ 1.1  & 12.2 $\pm$ 1.4  & 6.6 $\pm$ 1.0  & 21.8 $\pm$ 1.8 \\  
3         & 2.6 $\pm$ 0.8 & 5.2 $\pm$ 0.9  & 1.0 $\pm$ 0.4  & 16.5 $\pm$ 1.6   & 1.0 $\pm$ 0.4  &  7.2 $\pm$ 1.1  & 0.8 $\pm$ 0.3  & 21.0 $\pm$ 1.8 \\
4         & 0.00          & 2.3 $\pm$ 0.7  & 0.3 $\pm$ 0.2  &  4.3 $\pm$ 0.9   & 0.3 $\pm$ 0.2  &  2.9 $\pm$ 0.7  & 1.2 $\pm$ 0.4  &  0.8 $\pm$ 0.3 \\ 
$\langle$MLs$\rangle$ & 4.4 $\pm$ 0.4 & 3.2 $\pm$ 0.4  & 4.3 $\pm$ 0.1  & 12.0 $\pm$ 1.0   & 2.7 $\pm$ 0.7  &  6.9 $\pm$ 1.0  & 3.9 $\pm$ 0.8  & 15.0 $\pm$ 1.5 \\  \hline
\end{tabular}
\tablefoot{All values should be multiplied by 10$^{-3}$.  
The notation \ce{H2O}$^*$ and \ce{HOD}$^*$ denote photoexcitation of \ce{H2O} and \ce{HDO} in \ce{H2O} 
ice in which the H atom resulting from photodissociation kicks out a neigbouring \ce{H2O} molecule.  
The notation \ce{D2O}$^*$ and \ce{DOH}$^*$ indicate the analogous process in which the resulting D atom kicks 
out a neighbouring \ce{H2O} molecule.}
\label{Table3}
\end{table*}

\subsection{Ortho/para ratio of photodesorbed \ce{H2O}}
\label{orthopararatio}

Observations of water in space also provide constraints on the
ortho/para ratio of water in the gas. In diffuse interstellar clouds
and high temperature shocks, these measurements are consistent with a
ratio of 3, as expected from gas phase formation of water
\citep[e.g.,][]{Flagey13,Herczeg12,Emprechtinger13}.  In contrast, the
ortho/para ratio of water in protoplanetary disks and in
photon-dominated regions, such as the Orion Bar, is found to be much
lower, with values $< 1$ \citep{Hogerheijde11,Choi14}.  In these
regions, the water results from photodesorption of water ice, not from
thermal sublimation as is the case for comets \citep{Mumma11}.  The
question therefore arises to what extent the process of
photodesorption preserves the ortho/para ratio that was present in the
ice.

In the direct photodesorption mechanism, the H-OH bond is broken and
then reformed, so the ortho/para ratio is basically reset to the
statistical value of 3.  For the kick-out mechanism, however, the
original ortho/para ratio in the ice should be preserved.  The
relative importance of the direct and kick-out mechanisms depends on
the ice monolayer and to a lesser extent on the ice temperature.
Table~\ref{Table3} summarises the average values over the first four
monolayers of the two options. It is seen that the direct mechanism is
typically a factor 1.3 more efficient than the kick-out mechanism for
\ce{H2O}.  Thus, if the ortho/para ratio in the ice were low, say $\leq
0.5$, the gas-phase ortho/para ratio of the photodesorbed \ce{H2O}
would be significantly higher, at least 2, due to the photodesorption
process.

Low ortho/para ratios in the ice could occur if the ratio had
equilibrated to the grain temperature of typically 10--30 K, a process
that in itself is still poorly understood
\citep[e.g.,][]{Dulieu11}. The main message here is that due to the
partial reset during the photodesorption process, the observed gaseous
ortho/para ratio only partially reflects the original ice ratio.

\section{Conclusions} 
\label{conclusions}

This paper has investigated the deuterium fractionation that can occur
when OD, \ce{HDO} or \ce{D2O} molecules are liberated from a \ce{H2O}-rich ice
following a photodesorption event at a microscopic level. 
The results presented in Sect.~\ref{results} lead to the following conclusions:

\begin{enumerate}
\item No isotope fractionation occurs in the photodesorption of 
\ce{HDO} and \ce{H2O} from a mixed ice, even though kick-out by D is more efficient.  
The ratio of photodesorbed \ce{HDO} and 
\ce{H2O} is equal to the ratio of \ce{HDO} and \ce{H2O} in the ice.

\item No isotope fractionation occurs in the photodesorption of OD and 
OH upon \ce{HDO} photodissociation in \ce{H2O} ice.  The ratio of 
photodesorbed OD and OH is equal to the \ce{HDO}/\ce{H2O} ratio in 
the ice within 20\%.

\item The ratio of photodesorbed H and D will be more than twice the 
ratio of H and D present in the ice in HDO and \ce{H2O}, because 
photo-excitation of \ce{HDO} leads preferentially to desorption of H.  
Therefore, given enough time and enough photons photoprocessing 
of the outer layers of icy mantles could lead to 
enrichment in D relative to H.
\end{enumerate}

A possible mechanism for an increased ratio of \ce{HDO}/\ce{H2O} in
the gas phase compared with that initially in the ice may therefore be
prolonged photoprocessing of the outer layers of icy mantles of dust
particles.  Photo-excitation of \ce{HDO} leads preferentially to
desorption of H.  As a consequence, the ratio of photodesorbed H and D
will be greater than twice the ratio of H and D present in the ice.
If photoprocessing occurs over a long enough time, the outer layers
will develop a larger fraction of D, and, through grain surface
reactions of D with OH, therefore also of \ce{HDO}.  So ultimately,
the ratio of photodesorbed \ce{HDO} and \ce{H2O} can become enhanced
due to this indirect process.  Full gas-grain models that take the
multi-layer structure of the ice into account are needed to
investigate whether this is a plausible mechanism for isotope
fractionation of \ce{HDO}, since pure gas-phase processes can 
also lead to fractionation \citep[e.g.,][]{Aikawa12}.  If it is,
then the relative abundance of \ce{HDO} may also reflect the total
fluence of UV radiation (product of UV flux and time) that the outer
layers of icy mantles have been exposed to in the environment where
they are observed.

Finally, we note that the ortho/para ratio of water in the gas does
not directly reflect that in the ice but is partially reset by the
photodesorption process.

\begin{acknowledgements}
The authors are grateful to M.~C.~van Hemert for discussions on the
setup of the Wigner distributions and the data for the gas-phase
partial absorption spectra of HDO, and to S. Andersson for providing the
initial MD photodissociation code. This project was funded by NWO
astrochemistry grant No.~648.000.010 and 
from the European Union A-ERC grant 291141 CHEMPLAN.
\end{acknowledgements}

\Online

\appendix

\section{Auxiliary tables}
\label{auxiliarytables}

Tables~\ref{Table4}, \ref{Table5}, and \ref{Table6} contain the probabilities for X atom desorption, 
OX desorption, and \ce{X2O} and \ce{HDO} desorption (X=H,D), respectively,  at $T_\mathrm{ice}$~= 20, 30, and 60~K. 

\begin{table*}[!t]
\centering
\caption{X atom photodesorption probabilities at $T_\mathrm{ice}$~=~20~K, 30~K, and 60~K resulting from photoexcitation 
of a \ce{X2O} (X=H,D) or XOY (HOD or DOH) molecule present in a specific monolayer of \ce{H2O} ice.}
\begin{tabular}{ccccc} 
\hline\hline
ML        & H$_\mathrm{des}$/\ce{H2O}$^*$    & D$_\mathrm{des}$/\ce{D2O}$^*$    & H$_\mathrm{des}$/HOD$^*$         & D$_\mathrm{des}$/DOH$^*$         \\ \hline
\multicolumn{5}{c}{$T_\mathrm{ice}$~=~20~K} \\ \hline
1         & 0.865 $\pm$ 4.0$\times$10$^{-3}$ & 0.898 $\pm$ 3.9$\times$10$^{-3}$ & 0.913 $\pm$ 3.6$\times$10$^{-3}$ & 0.907 $\pm$ 3.7$\times$10$^{-3}$ \\
2         & 0.741 $\pm$ 5.9$\times$10$^{-3}$ & 0.665 $\pm$ 6.1$\times$10$^{-3}$ & 0.661 $\pm$ 6.0$\times$10$^{-3}$ & 0.626 $\pm$ 6.0$\times$10$^{-3}$ \\  
3         & 0.422 $\pm$ 6.1$\times$10$^{-3}$ & 0.401 $\pm$ 6.3$\times$10$^{-3}$ & 0.472 $\pm$ 6.3$\times$10$^{-3}$ & 0.419 $\pm$ 6.0$\times$10$^{-3}$ \\
4         & 0.261 $\pm$ 4.9$\times$10$^{-3}$ & 0.190 $\pm$ 5.1$\times$10$^{-3}$ & 0.194 $\pm$ 5.0$\times$10$^{-3}$ & 0.137 $\pm$ 4.2$\times$10$^{-3}$ \\ 
$\langle$MLs$\rangle$ & 0.572 $\pm$ 3.2$\times$10$^{-3}$ & 0.538 $\pm$ 3.2$\times$10$^{-3}$ & 0.559 $\pm$ 5.3$\times$10$^{-3}$ & 0.527 $\pm$ 5.0$\times$10$^{-3}$ \\ \hline 
\multicolumn{5}{c}{$T_\mathrm{ice}$~=~30~K} \\ \hline
1         & 0.878 $\pm$ 4.3$\times$10$^{-3}$ & 0.871 $\pm$ 4.3$\times$10$^{-3}$ & 0.863 $\pm$ 4.3$\times$10$^{-3}$ & 0.864 $\pm$ 4.3$\times$10$^{-3}$ \\
2         & 0.740 $\pm$ 5.9$\times$10$^{-3}$ & 0.670 $\pm$ 6.1$\times$10$^{-3}$ & 0.711 $\pm$ 5.7$\times$10$^{-3}$ & 0.655 $\pm$ 5.8$\times$10$^{-3}$ \\  
3         & 0.435 $\pm$ 6.3$\times$10$^{-3}$ & 0.452 $\pm$ 6.4$\times$10$^{-3}$ & 0.507 $\pm$ 6.3$\times$10$^{-3}$ & 0.453 $\pm$ 6.0$\times$10$^{-3}$ \\
4         & 0.319 $\pm$ 5.2$\times$10$^{-3}$ & 0.240 $\pm$ 5.6$\times$10$^{-3}$ & 0.310 $\pm$ 5.8$\times$10$^{-3}$ & 0.236 $\pm$ 5.1$\times$10$^{-3}$ \\ 
$\langle$MLs$\rangle$ & 0.593 $\pm$ 3.2$\times$10$^{-3}$ & 0.558 $\pm$ 3.2$\times$10$^{-3}$ & 0.598 $\pm$ 5.6$\times$10$^{-3}$ & 0.552 $\pm$ 5.0$\times$10$^{-3}$ \\ \hline 
\multicolumn{5}{c}{$T_\mathrm{ice}$~=~60~K} \\ \hline
1         & 0.872 $\pm$ 4.3$\times$10$^{-3}$ & 0.872 $\pm$ 4.3$\times$10$^{-3}$ & 0.904 $\pm$ 3.7$\times$10$^{-3}$ & 0.883 $\pm$ 4.0$\times$10$^{-3}$ \\
2         & 0.711 $\pm$ 6.0$\times$10$^{-3}$ & 0.706 $\pm$ 5.9$\times$10$^{-3}$ & 0.696 $\pm$ 5.8$\times$10$^{-3}$ & 0.667 $\pm$ 5.8$\times$10$^{-3}$ \\  
3         & 0.404 $\pm$ 6.3$\times$10$^{-3}$ & 0.350 $\pm$ 6.2$\times$10$^{-3}$ & 0.376 $\pm$ 6.1$\times$10$^{-3}$ & 0.298 $\pm$ 5.6$\times$10$^{-3}$ \\
4         & 0.302 $\pm$ 5.9$\times$10$^{-3}$ & 0.230 $\pm$ 5.4$\times$10$^{-3}$ & 0.306 $\pm$ 5.7$\times$10$^{-3}$ & 0.204 $\pm$ 4.8$\times$10$^{-3}$ \\ 
$\langle$MLs$\rangle$ & 0.572 $\pm$ 3.2$\times$10$^{-3}$ & 0.539 $\pm$ 3.2$\times$10$^{-3}$ & 0.570 $\pm$ 5.4$\times$10$^{-3}$ & 0.513 $\pm$ 5.0$\times$10$^{-3}$ \\ \hline
\end{tabular}
\tablefoot{ML1 is the top monolayer and $\langle$MLs$\rangle$ denotes the average over the top four monolayers.}
\label{Table4}
\end{table*}

\begin{table*}[!t]
\centering
\caption{OX radical photodesorption probabilities at $T_\mathrm{ice}$~=~20~K, 30~K, and 60~K resulting from photoexcitation 
of a \ce{X2O} (X=H,D) or XOY (HOD or DOH) molecule present in a specific monolayer of \ce{H2O} ice.}
\begin{tabular} {ccccc} 
\hline\hline
ML        & OH$_\mathrm{des}$/\ce{H2O}$^*$   & OD$_\mathrm{des}$/\ce{D2O}$^*$ & OD$_\mathrm{des}$/HOD$^*$         & OH$_\mathrm{des}$/DOH$^*$        \\ \hline
\multicolumn{5}{c}{$T_\mathrm{ice}$~=~20~K} \\ \hline
1         & 0.026 $\pm$ 2.7$\times$10$^{-3}$ & 0.043 $\pm$ 2.6$\times$10$^{-3}$ & 0.028 $\pm$ 2.1$\times$10$^{-3}$  & 0.055 $\pm$ 2.9$\times$10$^{-3}$ \\
2         & 0.021 $\pm$ 2.7$\times$10$^{-3}$ & 0.038 $\pm$ 2.5$\times$10$^{-3}$ & 0.022 $\pm$ 1.8$\times$10$^{-3}$  & 0.049 $\pm$ 2.7$\times$10$^{-3}$ \\  
3         & (1.0 $\pm$ 0.3)$\times$10$^{-3}$ & (1.7 $\pm$ 0.5)$\times$10$^{-3}$ & (4.7 $\pm$ 2.7)$\times$10$^{-4}$  & (1.7 $\pm$ 0.5)$\times$10$^{-3}$ \\
4         & 0.00                             & 0.00                             & 0.00                              & 0.00 \\
$\langle$MLs$\rangle$ & 0.012 $\pm$ 7.0$\times$10$^{-4}$ & 0.021$\pm$ 9.1$\times$10$^{-4}$  & 0.012 $\pm$ 1.4$\times$10$^{-3}$  & 0.026 $\pm$ 2.0$\times$10$^{-3}$ \\  \hline 
\multicolumn{5}{c}{$T_\mathrm{ice}$~=~30~K} \\ \hline
1         & 0.047 $\pm$ 2.9$\times$10$^{-3}$ & 0.069 $\pm$ 3.3$\times$10$^{-3}$ & 0.029  $\pm$ 2.1$\times$10$^{-3}$ & 0.054 $\pm$ 2.8$\times$10$^{-3}$ \\
2         & 0.011 $\pm$ 1.8$\times$10$^{-3}$ & 0.019 $\pm$ 1.8$\times$10$^{-3}$ & (5.5 $\pm$ 0.9)$\times$10$^{-3}$  & 0.015 $\pm$ 1.5$\times$10$^{-3}$ \\  
3         & (1.7 $\pm$ 0.5)$\times$10$^{-4}$ & 0.012 $\pm$ 1.4$\times$10$^{-3}$ & (3.8 $\pm$ 0.8)$\times$10$^{-3}$  & (6.8 $\pm$ 0.9)$\times$10$^{-3}$ \\
4         & 0.00                             & (1.7 $\pm$ 1.7)$\times$10$^{-4}$ & (2.0 $\pm$ 2.0)$\times$10$^{-4}$  & 0.00                             \\ 
$\langle$MLs$\rangle$ & 0.014 $\pm$ 7.5$\times$10$^{-4}$ &0.025 $\pm$ 1.0$\times$10$^{-3}$  & (9.5 $\pm$ 1.2)$\times$10$^{-3}$  & 0.019 $\pm$ 2.0$\times$10$^{-3}$ \\  \hline 
\multicolumn{5}{c}{$T_\mathrm{ice}$~=~60~K} \\ \hline 
1         & 0.029 $\pm$ 2.2$\times$10$^{-3}$ & 0.053 $\pm$ 2.9$\times$10$^{-3}$ & 0.013 $\pm$ 1.4$\times$10$^{-3}$  & 0.050 $\pm$ 2.7$\times$10$^{-3}$ \\
2         & 0.012 $\pm$ 1.4$\times$10$^{-3}$ & 0.026 $\pm$ 2.0$\times$10$^{-3}$ & (7.0 $\pm$ 1.0)$\times$10$^{-3}$  & 0.020 $\pm$ 1.7$\times$10$^{-3}$ \\  
3         & (2.3 $\pm$ 0.6)$\times$10$^{-3}$ & (6.0 $\pm$ 0.1)$\times$10$^{-3}$ & (3.5 $\pm$ 0.8)$\times$10$^{-3}$  & 0.012 $\pm$ 1.3$\times$10$^{-3}$ \\
4         & 0.00                             & 0.00                             & 0.00                              & 0.00                             \\ 
$\langle$MLs$\rangle$ & 0.011$\pm$ 6.7$\times$10$^{-4}$  & 0.021 $\pm$ 9.3$\times$10$^{-4}$ & (5.9 $\pm$ 1.0)$\times$10$^{-3}$  & 0.020 $\pm$ 2.0$\times$10$^{-3}$ \\   \hline
\end{tabular}
\tablefoot{ML1 is the top monolayer and $\langle$MLs$\rangle$ denotes the average over the top four monolayers.}
\label{Table5}
\end{table*}


\begin{table*}[!t]
\centering
\footnotesize
\caption{Total \ce{X2O} (X=H,D) or XOY (HOD or DOH) photodesorption probabilities at $T_\mathrm{ice}$~=~20~K, 30~K, and 60~K per monolayer due to the direct and the 
kick-out mechanism for \ce{X2O} and \ce{XOY} photodissociation in \ce{H2O} ice.}
\begin{tabular}{ccccccccc}
\hline \hline
   & \multicolumn{2}{c}{\ce{H2O}$^*$} & \multicolumn{2}{c}{\ce{D2O}$^*$} & \multicolumn{2}{c}{HOD$^*$} & \multicolumn{2}{c}{DOH$^*$} \\
ML & \ce{H2O}$_\mathrm{direct}$ & \ce{H2O}$_\mathrm{kicked}$  & \ce{D2O}$_\mathrm{direct}$ & \ce{H2O}$_\mathrm{kicked}$ & HOD$_\mathrm{direct}$ & 
\ce{H2O}$_\mathrm{kicked}$ & DOH$_\mathrm{direct}$ & \ce{H2O}$_\mathrm{kicked}$  \\ \hline
\multicolumn{9}{c}{$T_\mathrm{ice}$~=~20~K} \\ \hline
1         & 7.8 $\pm$ 1.3 & 0.7 $\pm$ 0.4 & 3.3 $\pm$ 0.7 &  3.3 $\pm$ 0.7 & 4.5 $\pm$ 0.9  & 1.6 $\pm$ 0.5  & 3.7 $\pm$ 0.8 &  4.2 $\pm$ 0.8 \\
2         & 5.5 $\pm$ 1.2 & 2.8 $\pm$ 0.9 & 7.1 $\pm$ 1.1 & 12.5 $\pm$ 1.4 & 7.1 $\pm$ 1.1  & 3.7 $\pm$ 0.8  & 5.9 $\pm$ 0.9 &  4.2 $\pm$ 0.8 \\  
3         & 0.3 $\pm$ 0.2 & 5.9 $\pm$ 1.1 & 1.7 $\pm$ 0.5 & 13.0 $\pm$ 1.5 & 1.0 $\pm$ 0.4  & 0.8 $\pm$ 0.4  & 0.3 $\pm$ 0.2 &  7.0 $\pm$ 1.0 \\
4         & 0.2 $\pm$ 0.2 & 0.2 $\pm$ 0.2 & 0.00          & 14.0 $\pm$ 1.5 & 0.2 $\pm$ 0.2  & 3.9 $\pm$ 0.8  & 0.00          & 16.0 $\pm$ 1.5 \\ 
$\langle$MLs$\rangle$ & 3.4 $\pm$ 0.4 & 2.4 $\pm$ 0.3 & 3.0 $\pm$ 0.4 & 11.0 $\pm$ 0.7 & 3.2 $\pm$ 0.7  & 2.5 $\pm$ 0.6  & 2.5 $\pm$ 0.6 &  7.9 $\pm$ 1.1 \\  \hline 
\multicolumn{9}{c}{$T_\mathrm{ice}$~=~30~K} \\ \hline
1         & 2.6 $\pm$ 0.9 &  4.2 $\pm$ 0.9 & 4.5 $\pm$ 0.9 &  0.5 $\pm$ 0.3 & 4.6 $\pm$ 0.9  & 1.0 $\pm$ 0.4 & 5.6 $\pm$ 0.9 &  0.2 $\pm$ 0.2 \\
2         & 2.9 $\pm$ 1.1 &  5.7 $\pm$ 0.9 & 4.8 $\pm$ 0.9 & 17.0 $\pm$ 0.2 & 3.6 $\pm$ 0.8  & 2.9 $\pm$ 0.7 & 7.5 $\pm$ 0.1 & 14.4 $\pm$ 1.5 \\  
3         & 1.7 $\pm$ 1.2 & 10.0 $\pm$ 0.2 & 2.2 $\pm$ 0.6 & 10.6 $\pm$ 1.3 & 1.4 $\pm$ 0.5  & 2.8 $\pm$ 0.7 & 2.8 $\pm$ 0.6 &  5.2 $\pm$ 0.9 \\
4         & 0.9 $\pm$ 0.5 &  3.1 $\pm$ 0.8 & 0.3 $\pm$ 0.2 &  3.8 $\pm$ 0.8 & 0.6 $\pm$ 0.3  & 2.5 $\pm$ 0.6 & 0.00          &  5.5 $\pm$ 0.9 \\ 
$\langle$MLs$\rangle$ & 2.0 $\pm$ 0.3 &  5.8 $\pm$ 0.3 & 3.0 $\pm$ 0.4 &  8.0 $\pm$ 0.6 & 2.5 $\pm$ 0.6  & 2.3 $\pm$ 0.6 & 4.0 $\pm$ 0.8 &  6.3 $\pm$ 1.0 \\  \hline 
\multicolumn{9}{c}{$T_\mathrm{ice}$~=~60~K} \\ \hline
1         & 8.8 $\pm$ 1.0 & 4.8 $\pm$ 0.9  & 5.3 $\pm$ 0.9 & 11.0 $\pm$ 1.3 & 3.9 $\pm$ 0.8 & 10.0 $\pm$ 1.3  & 3.1 $\pm$ 0.7 & 21.5 $\pm$ 1.8 \\
2         & 2.7 $\pm$ 1.2 & 4.7 $\pm$ 0.9  & 3.2 $\pm$ 0.7 & 20.0 $\pm$ 1.8 & 3.9 $\pm$ 0.8 & 10.8 $\pm$ 1.3  & 4.6 $\pm$ 0.8 & 15.4 $\pm$ 1.5 \\  
3         & 1.5 $\pm$ 0.9 & 7.3 $\pm$ 1.2  & 3.0 $\pm$ 0.7 & 23.5 $\pm$ 1.9 & 1.1 $\pm$ 0.4 &  2.4 $\pm$ 0.6  & 4.1 $\pm$ 0.8 &  9.2 $\pm$ 1.1 \\
4         & 0.00          & 0.8 $\pm$ 0.5  & 0.3 $\pm$ 0.2 &  4.3 $\pm$ 0.9 & 0.00          &  1.6 $\pm$ 0.5  & 0.00          &  0.9 $\pm$ 0.4 \\ 
$\langle$MLs$\rangle$ & 3.3 $\pm$ 0.4 & 4.4 $\pm$ 0.4  & 2.9 $\pm$ 0.3 & 14.0 $\pm$ 0.8 & 2.2 $\pm$ 0.6 &  6.2 $\pm$ 1.0  & 2.9 $\pm$ 0.7 & 12.0 $\pm$ 1.3 \\ \hline
\end{tabular}
\tablefoot{All values should be multiplied by 10$^{-3}$.  
The notation \ce{H2O}$^*$ and \ce{HOD}$^*$ denote photoexcitation of \ce{H2O} and \ce{HDO} in \ce{H2O} 
ice in which the H atom resulting from photodissociation kicks out a neigbouring \ce{H2O} molecule.  
The notation \ce{D2O}$^*$ and \ce{DOH}$^*$ indicate the analogous process in which the resulting D atom kicks 
out a neighbouring \ce{H2O} molecule.}
\label{Table6}
\end{table*}

\section{Fitting formulae for photodesorption probabilities}
\label{fittingformula}

Tables~\ref{Table1}, \ref{Table2}, \ref{Table3}, \ref{Table4}, \ref{Table5}, and \ref{Table6} 
list the {\em total} probabilities for X desorption, 
OX desorption, and \ce{X2O} and \ce{HDO} desorption (X=H,D) following a dissociation event, as a function 
of both monolayer and ice temperature.  
These tables also give the average probabilities, over the top four monolayers, for each 
species.  
For use in astrochemical models, it is useful to know the probability 
(per monolayer) of {\em every} potential outcome, rather than the total probability for the desorption of each species.  
This is because, in full gas-grain models, one is also interested in the composition of the 
ice mantle, as well as the gas.  

As discussed in the main body of this paper, there are six potential outcomes following a dissociation 
event which can lead to a change in composition of both the ice and gas.  
For example, for HDO which is dissociated into H~+~OD, 
\begin{align}
\ce{HDO} + h\nu & \longrightarrow \ce{H}_\mathrm{des} + \ce{OD}_\mathrm{trapped}     \label{eqnB1}\\ 
                & \longrightarrow \ce{H}_\mathrm{trapped} + \ce{OD}_\mathrm{des}     \label{eqnB2}\\
                & \longrightarrow \ce{H}_\mathrm{des} + \ce{OD}_\mathrm{des}         \label{eqnB3}\\
                & \longrightarrow \ce{H}_\mathrm{trapped} + \ce{OD}_\mathrm{trapped} \label{eqnB4}\\
                & \longrightarrow \ce{HDO}_\mathrm{direct}                           \label{eqnB5}\\
                & \longrightarrow \ce{H2O}_\mathrm{kicked}. \label{eqnB6}                           
\end{align}
Eq.~\ref{eqnB6} is the process known as ``kick out'' whereby a neighbouring 
\ce{H2O} is ejected from the ice via momentum transfer from an excited photofragment.  
The probabilities of each of these events as a function of monolayer and ice temperature 
have been compiled from the raw data of the molecular dynamics simulations and are available in electronic form only.  
There is a seventh possibility in which the photofragments recombine to reform 
\ce{HDO} which remains trapped in the ice.  
This process does not change the gas or ice composition and thus we have not listed 
the probabilities for this outcome here; however, these data are necessary if one 
is interested in extrapolating the probabilities to deeper monolayers, ML~$>$~4.  

To determine the desorption probabilities at temperatures and in monolayers outside of those tabulated, 
one can simply interpolate/extrapolate using, for example, cubic spline interpolation.  
However, when extrapolating to determine probabilities for deeper monolayers, ML~$>$~4, 
one should take care to ensure that, deep into the ice mantle, the probabilities 
for outcomes \ref{eqnB1}, \ref{eqnB2}, \ref{eqnB3}, 
\ref{eqnB5}, and \ref{eqnB6} tend to 0, and the probability for outcome \ref{eqnB4} 
tends to $1 - P_\mathrm{recom}$, where $P_\mathrm{recom}$ is the probability that the 
photofragments recombine to reform the molecule (which remains trapped in the ice).  
Deeper into the ice, desorption events become increasingly less probable and the most probable 
outcome becomes trapping of the photofragments (or the reformed molecule, following recombination).  
In addition, at very low coverage, ML~$<$~1, the rates for all outcomes should tend to  
0 as ML~$\to$~0.  

\begin{figure*}[!t]
\subfigure{\includegraphics[width=0.5\textwidth]{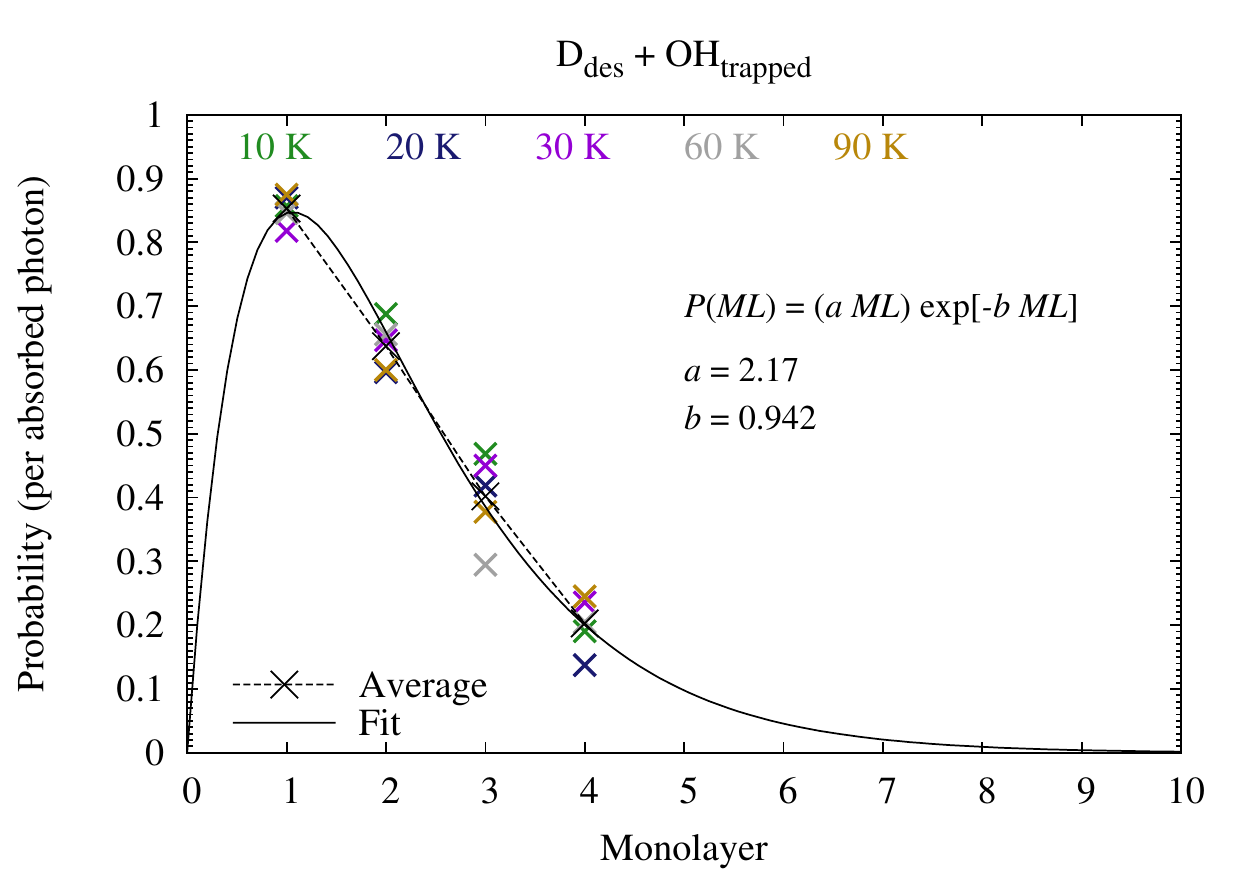}}
\subfigure{\includegraphics[width=0.5\textwidth]{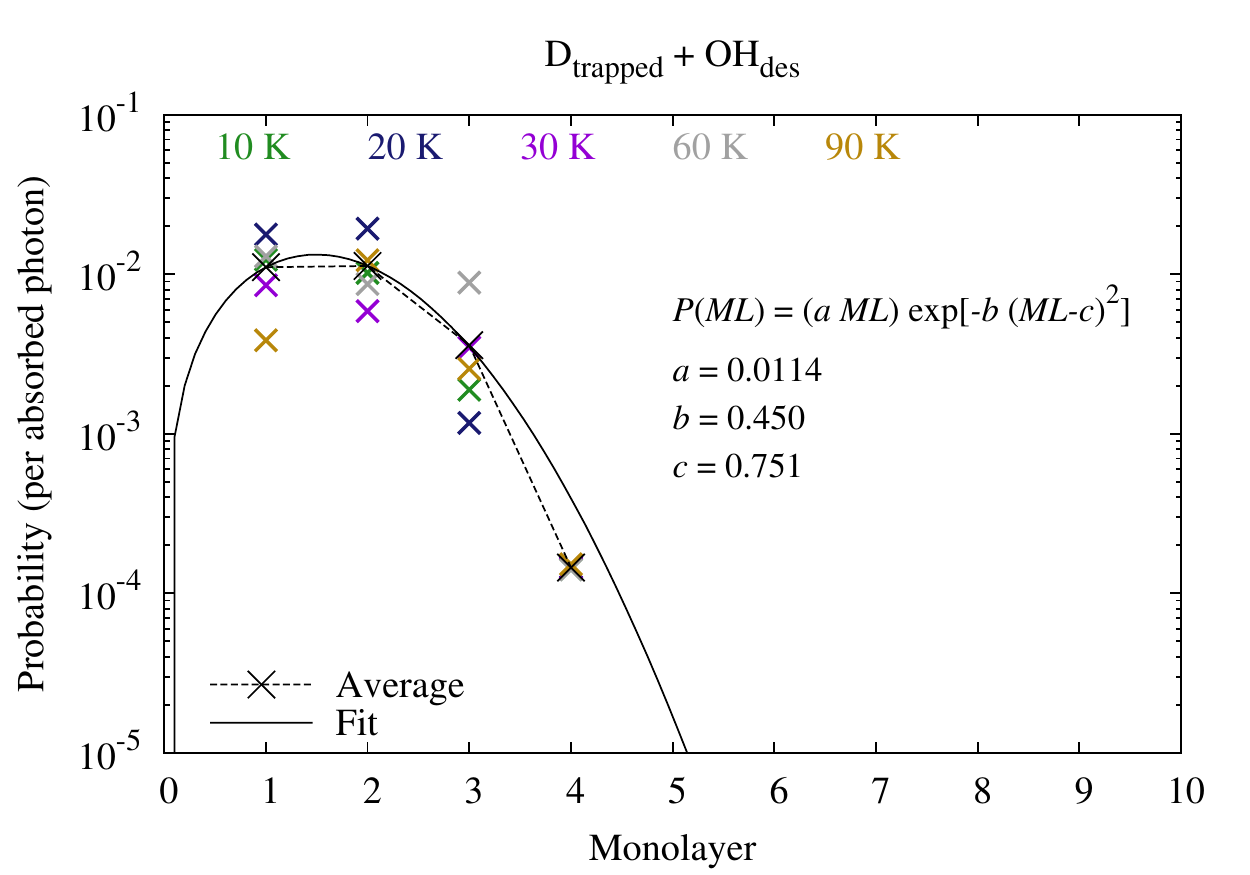}}
\subfigure{\includegraphics[width=0.5\textwidth]{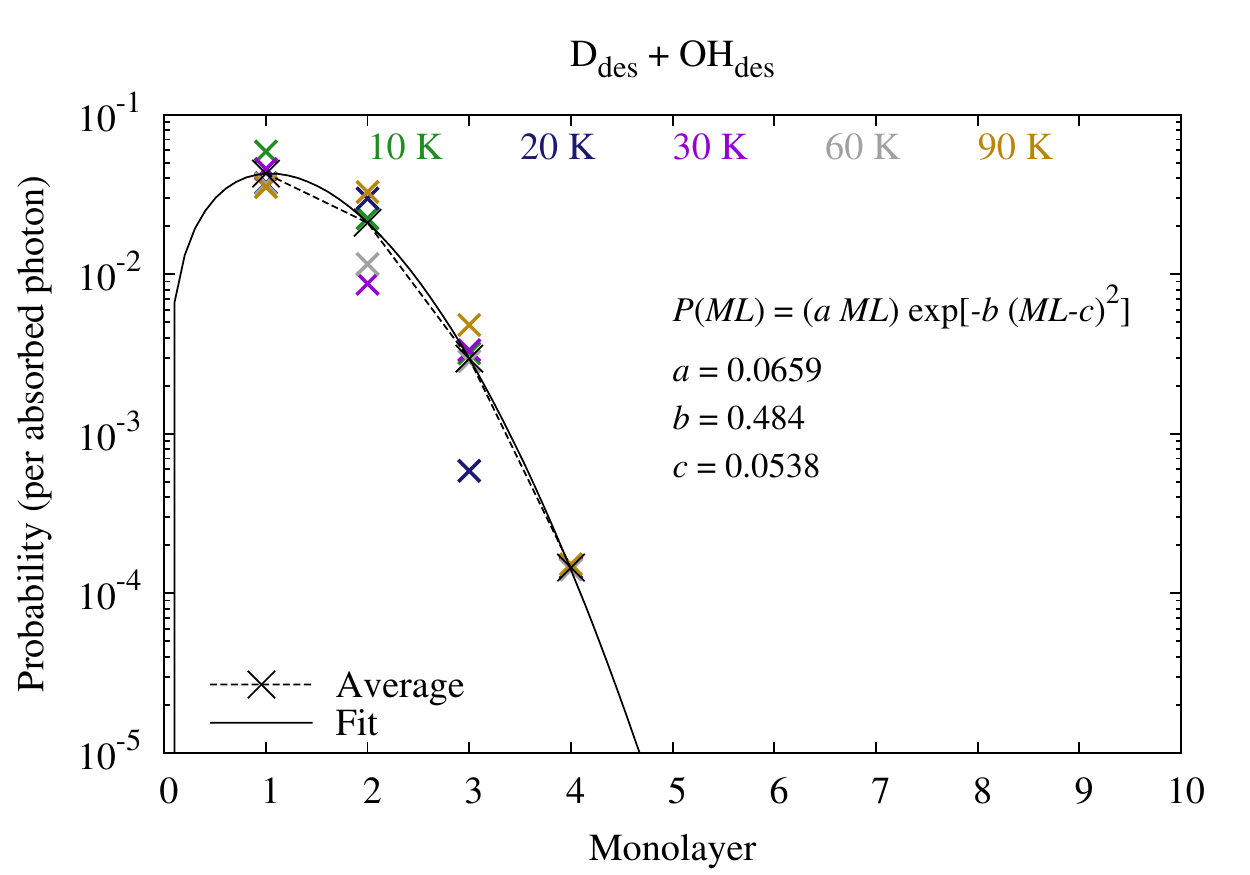}}
\subfigure{\includegraphics[width=0.5\textwidth]{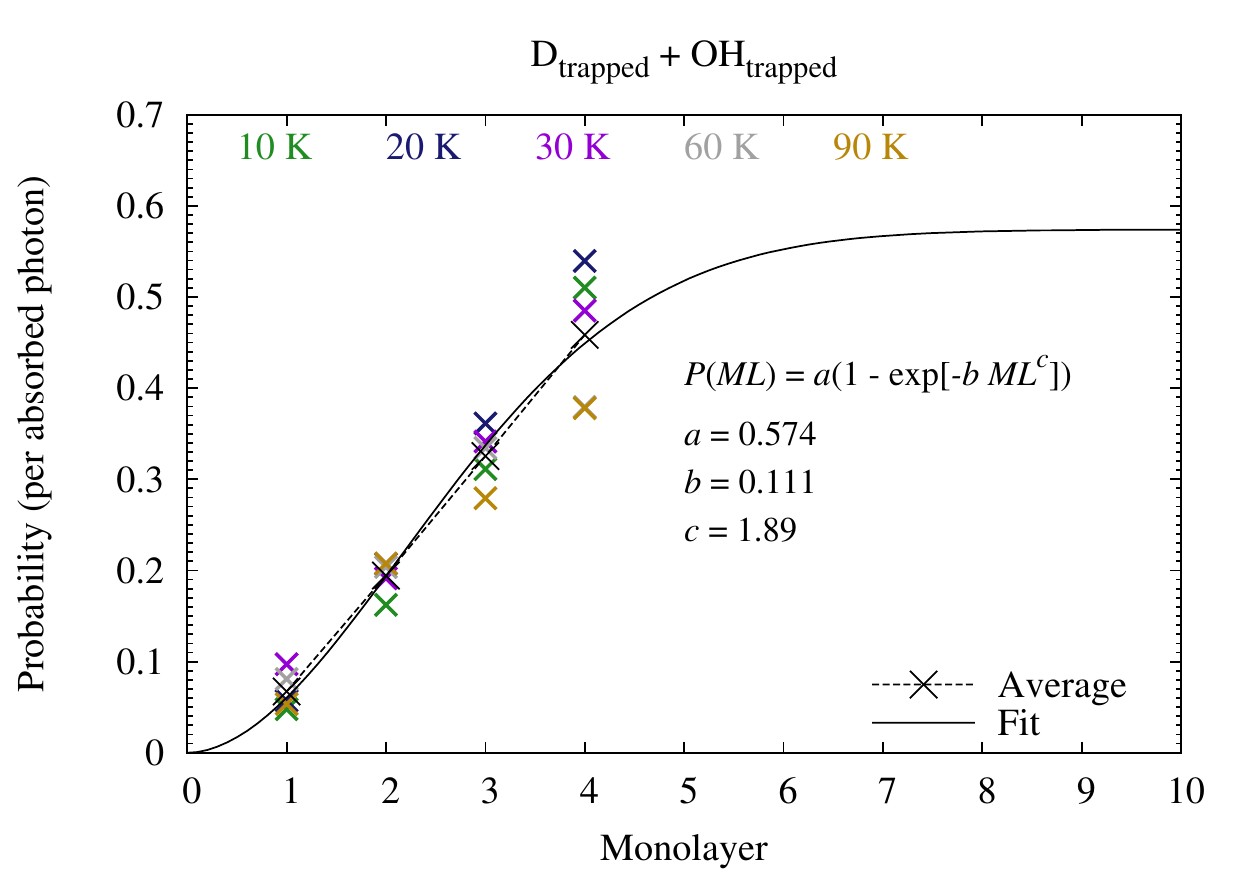}}
\subfigure{\includegraphics[width=0.5\textwidth]{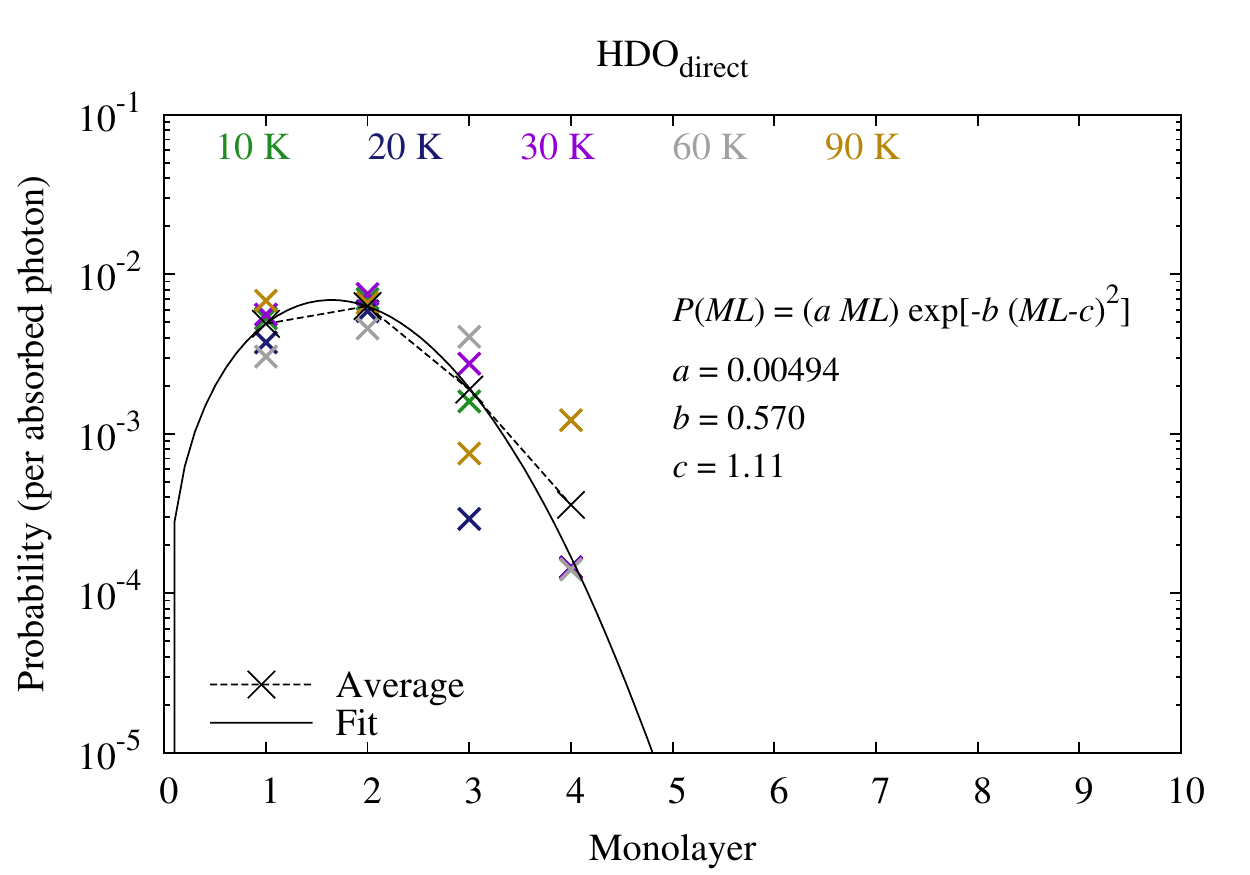}}
\subfigure{\includegraphics[width=0.5\textwidth]{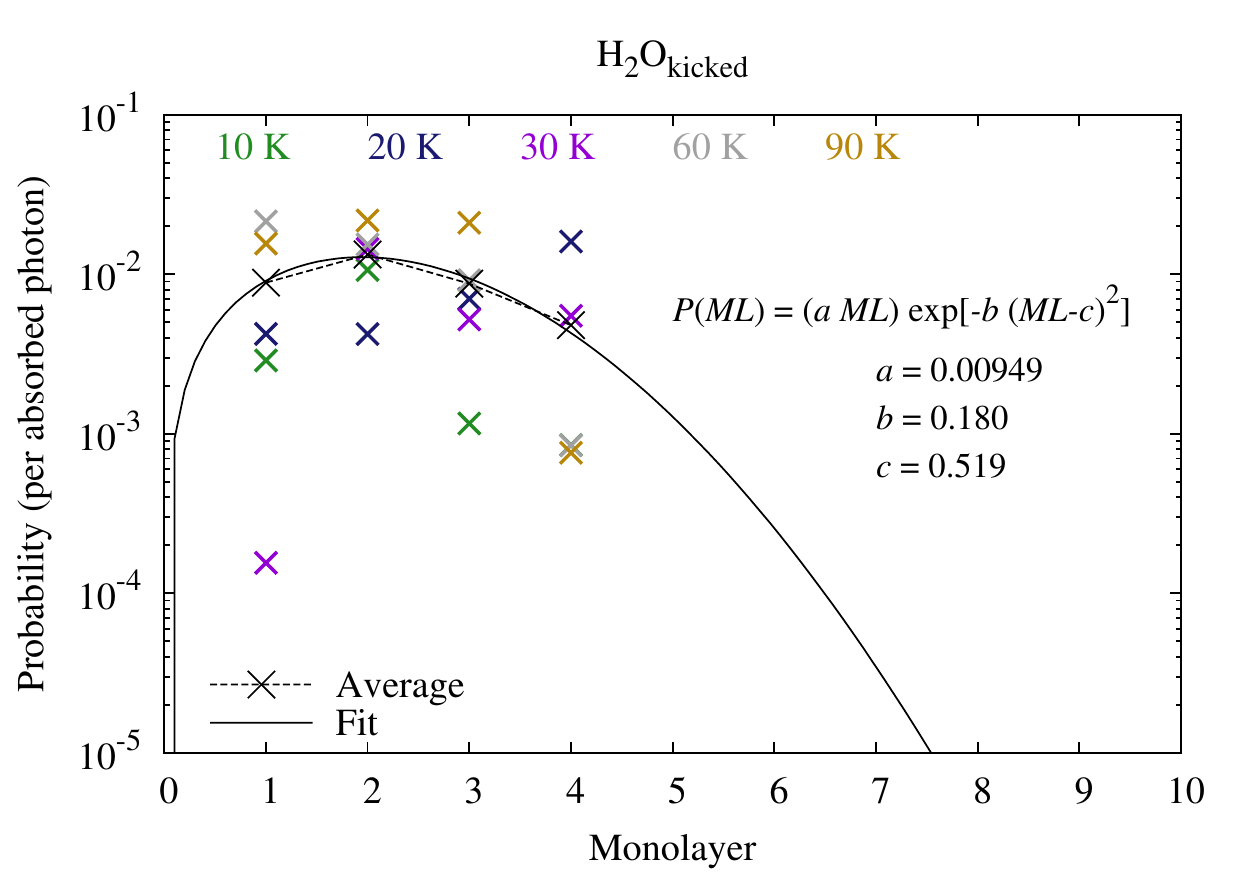}}
\caption{Temperature-specific probabilities, temperature-averaged probabilities, 
and fitted functions for each outcome as a function monolayer for 
HDO photodissociation into D~+~OH.}
\label{FigureB1}
\end{figure*}

In Table~\ref{TableFits} we present our fitting functions and corresponding 
best-fit parameters for the {\em temperature-averaged} probabilities per 
monolayer for each outcome.  
The probabities are well fitted using a Gaussian-like function with the exception 
of the outcomes leading to trapping of the OY radical for which an exponential-like  
function was found to be more appropriate for describing the asymptotic behaviour 
of the probabilities towards deeper monolayers ($\gg$~4). 
In Fig.~\ref{FigureB1} we present the probability per monolayer at each temperature 
and the temperature-averaged probabilities per monolayer along with the fitted 
functions for the example of DOH$^*$.  
The probabilities were fitted using the nonlinear least-squares (NLLS) 
Marquardt-Levenberg algorithm \citep[][]{Marquardt63}.  
The probabilities are a much stronger function of monolayer than 
temperature; hence, our decision to fit functions with respect to monolayer 
only.  

For implementation in chemical models which adopt the rate equation
method for describing the ice chemistry and gas-grain balance, the
probabilities per monolayer should be multiplied by the rate of
arrival of UV photons in the wavelength range 1650--1300 \AA\ onto the
grain surface times the absorption cross section of a UV photon by a
grain-surface site (or molecule, in this case, \ce{HDO}).  The total
desorption rate is then determined by integrating the desorption rate
per monolayer over the total number of monolayers on the grain.  The
probabilities can be directly employed in stochastic chemical models
in which the discrete nature of chemical reactions are taken into
account \citep[see, e.g., ][]{Cuppen07}.

\section{Photodesorption and fractionation}

This section investigates whether photodesorption ultimately 
also leads to fractionation of HDO/\ce{H2O} in the gas.  
We can estimate the total photodesorption probability ratio between \ce{HDO}
and \ce{H2O} by taking into account the direct and kicked out
mechanism in both cases.  The probability of \ce{HDO} photodesorption
through the direct mechanism is given by
\begin{equation}
\label{eq:GJ1}
P^\mathrm{dirdes}_\ce{HDO} = r_\ce{HDO} P^\mathrm{direct}(\ce{HDO}^*).
\end{equation}
In Eq.~\ref{eq:GJ1}, $P^\mathrm{direct}$(\ce{HDO}$^*$) is the probability
that upon photo-excitation of \ce{HDO} (the generic case) the \ce{HDO}
recombines and desorbs directly. It can be approximately calculated
using
\begin{equation}
\label{eq:GJ2}
P_\mathrm{direct}(\ce{HDO}^*) = \frac{2}{3}P_\mathrm{direct}(\ce{HOD}^*) +  \frac{1}{3}P_\mathrm{direct}(\ce{DOH}^*)
\end{equation}
and $r_\ce{HDO}$ is the original \ce{HDO}/\ce{H2O} ratio in the ice 
(of the order of 0.01 or less as indicated by observations). 
In Eq.~\ref{eq:GJ2}, the probabilities on the right hand
side are the probabilities for the direct mechanism for photodesorbing
\ce{HDO} averaged over the top four monolayers and presented in
Tables~\ref{Table3} and \ref{Table6}.

The probability of \ce{H2O} photodesorption through the direct
mechanism is given by 
\begin{equation}
\label{eq:GJ3}
P^\mathrm{dirdes}_\ce{H2O} = (1- r_\ce{HDO})P_\mathrm{direct}(\ce{H2O}^*)
\end{equation}
In Eq.~\ref{eq:GJ3}, $P_\mathrm{direct}$(\ce{H2O}$^*$) is the
probability that upon photo-excitation \ce{H2O} recombines and desorbs
directly.  It can obtained directly from Tables~\ref{Table3} and
\ref{Table6}.

As can be seen from Table~\ref{Table1} and after using
Eq.~\ref{eq:GJ2}, $P_\mathrm{direct}$(\ce{HDO}$^*$) and
$P_\mathrm{direct}$(\ce{H2O}$^*$) are roughly the same. 
As a result
\begin{equation}
\label{eq:GJ4}
P^\mathrm{dirdes}_\ce{HDO}/P^\mathrm{dirdes}_\ce{H2O} = r_\ce{HDO}/(1-r_\ce{HDO}) =  r_\ce{HDO}/r_\ce{H2O}
\end{equation}
meaning that there is no isotope fractionation due to the direct mechanism.

Now consider the kick-out mechanism. 
The indirect probabilities can be written as follows:
\begin{align}
P^\mathrm{KOdes}_\ce{HDO} & = r_\ce{HDO} \times r_\ce{HDO} \times P_\mathrm{KO}(\ce{HDO};\ce{HDO}^*) \nonumber \\
 &+ (1 - r_\ce{HDO}) \times r_\ce{HDO} \times P_\mathrm{KO}(\ce{HDO};\ce{H2O}^*) \label{eq:GJ5}
\end{align}
and
\begin{align}
P^\mathrm{KOdes}_\ce{H2O} & = (1 - r_\ce{HDO}) \times (1 - r_\ce{HDO}) \times P_\mathrm{KO}(\ce{H2O};\ce{H2O}^*) \nonumber \\
                          & + r_\ce{HDO}  \times  (1 - r_\ce{HDO}) \times P_\mathrm{KO}(\ce{H2O};\ce{HDO}^*) \label{eq:GJ6} 
\end{align}
In Eqs.~\ref{eq:GJ5} and \ref{eq:GJ6}, $P^\mathrm{KOdes}_\ce{HXO}$ is the
probability of desorption of HXO through the kick-out mechanism, where 
X is either H or D. 
Furthermore,
$P_\mathrm{KO}$(HX$^{1}$O;HX$^{2}$O$^*$) is the probability that
HX$^{1}$O is kicked out after photo-excitation of HX$^{2}$O, where
X$^{1}$ can either be H or D, and X$^{2}$ can also be H or D. 
As for the direct mechanism, we can approximately calculate
$P_\mathrm{KO}$(HX$^{1}$O;HX$^{2}$O$^*$) from
\begin{align}
P_\mathrm{KO}(\mathrm{HX^{1}O};\mathrm{HDO}^*) & = \frac{2}{3}P_\mathrm{KO}(\mathrm{HX^{1}O};\mathrm{HOD}^*) \nonumber \\
                                               & + \frac{1}{3}P_\mathrm{KO}(\mathrm{HX^{1}O};\mathrm{DOH}^*) \label{eq:GJ7}
\end{align}
The two quantities on the right hand side of Eq.~\ref{eq:GJ7} have been tabulated for 
X$^{1}$ equal to H in Tables~\ref{Table3} and \ref{Table6}.

Because we have only calculated probabilities that \ce{H2O} is kicked out, we make the following approximations,
\begin{equation}
\label{eq:GJ8}
P_\mathrm{KO}(\ce{HDO};\ce{HOD}^*) = P_\mathrm{KO}(\ce{H2O};\ce{HOD}^*)
\end{equation}
\begin{equation}
\label{eq:GJ9}
P_\mathrm{KO}(\ce{HDO};\ce{DOH}^*) = P_\mathrm{KO}(\ce{H2O};\ce{DOH}^*)
\end{equation}
\begin{equation}
\label{eq:GJ10}
P_\mathrm{KO}(\ce{HDO};\ce{HDO}^*) = P_\mathrm{KO}(\ce{H2O};\ce{HDO}^*)
\end{equation}
The right hand values of Eqs.~\ref{eq:GJ8} and \ref{eq:GJ9} can be
directly obtained from Tables~\ref{Table3} and \ref{Table6}.
$P_\mathrm{KO}$(\ce{HDO};\ce{HDO}$^*$) can be computed using the approximation in
Eq.~\ref{eq:GJ10} and using Eq.~\ref{eq:GJ7} and Tables~\ref{Table3}
and \ref{Table6}.

Using Eqs.~\ref{eq:GJ8}--\ref{eq:GJ10}, Eq.~\ref{eq:GJ5} can be
rewritten as
\begin{align}
P^\mathrm{KOdes}_\ce{HDO} & \approx r_\ce{HDO} \times r_\ce{HDO} \times P_\mathrm{KO}(\ce{H2O};\ce{HDO}^*) \nonumber\\
                          & + (1 - r_\ce{HDO}) \times r_\ce{HDO} \times  P_\mathrm{KO}(\ce{H2O};\ce{H2O}^*) \label{eq:GJ12}
\end{align}
Most importantly, for 10 and 20~K we have approximately that (see Tables~\ref{Table3} and \ref{Table6})
\begin{equation}
\label{eq:GJ13}
P_\mathrm{KO}(\ce{H2O};\ce{HDO}^*) \approx P_\mathrm{KO}(\ce{H2O};\ce{H2O}^*).
\end{equation}
Inserting Eq.~\ref{eq:GJ13} in Eq.~\ref{eq:GJ12} yields
\begin{equation}
\label{eq:GJ14}
P^\mathrm{KOdes}_\ce{HDO}  \approx r_\ce{HDO} \times P_\mathrm{KO}(\ce{H2O};\ce{H2O}^*)
\end{equation}
and inserting Eq.~\ref{eq:GJ13} in Eq.~\ref{eq:GJ6} yields
\begin{equation}
\label{eq:GJ15}
P^\mathrm{KOdes}_\ce{H2O} = (1 - r_\ce{HDO}) \times P_\mathrm{KO}(\ce{H2O};\ce{H2O}^*)
\end{equation}
From Eqs.~\ref{eq:GJ14} and \ref{eq:GJ15}, we can derive that
\begin{equation}
\label{eq:GJ16}
P^\mathrm{KOdes}_\ce{HDO}/P^\mathrm{KOdes}_\ce{H2O} = r_\ce{HDO}/r_\ce{H2O}
\end{equation}
meaning that there should be no isotope fractionation due to the
indirect kick-out mechanism. Taken together, Eqs.~\ref{eq:GJ4} and
\ref{eq:GJ16} ensure that the ratio of desorbed \ce{HDO} over desorbed
\ce{H2O} in the ice is given by
\begin{equation}
\label{eq:GJ17}
P^\mathrm{des}_\ce{HDO}/P^\mathrm{des}_\ce{H2O} = r_\ce{HDO}/r_\ce{H2O}
\end{equation}
which means that this ratio is simply equal to the ratio of \ce{HDO} and
\ce{H2O} in the ice. Therefore, isotope fractionation does not occur
for \ce{HDO} and \ce{H2O} photodesorption.

\end{document}